\documentclass{aa}
\usepackage{graphicx}
\def\iras{{\it IRAS}}
\def\solyr{M_\odot~{\rm yr}^{-1}}
\begin{document}
\title{Star formation rate in galaxies from UV, IR, and H$\alpha$
estimators}
\author{H. Hirashita \inst{1}\fnmsep\inst{2}
\fnmsep\thanks{Postdoctoral Fellow of the Japan Society for the
Promotion of Science (JSPS) for Research Abroad.
Present address: Graduate School of Science,
Nagoya University, Chikusa-ku, Nagoya 464-8602, Japan.},
V. Buat \inst{2}
         \and
A. K. Inoue \inst{3}
\fnmsep\thanks{Research Fellow of JSPS. Present address:
Department of Physics, Kyoto University, Sakyo-ku, Kyoto
606-8502, Japan.}
}
\offprints{H. Hirashita,\\
\email{hirashita@u.phys.nagoya-u.ac.jp}}
\institute{Osservatorio Astrofisico di Arcetri, Largo Enrico Fermi,
5, 50125 Firenze, Italy\\
\email{hirashita@u.phys.nagoya-u.ac.jp}
\and
Laboratoire d'Astrophysique de Marseille, BP 8, 13376 Marseille
Cedex 12, France\\
\email{veronique.buat@astrsp-mrs.fr}
        \and
Department of Astronomy, Faculty of Science, Kyoto University,
Sakyo-ku, Kyoto 606-8502, Japan\\
\email{akinoue@scphys.kyoto-u.ac.jp}
}
\date{Received 18 March 2003 / Accepted 16 July 2003}
\abstract{
Infrared (IR) luminosity of galaxies originating from dust
thermal emission can be used as an indicator of the star
formation rate (SFR). Inoue et al.\ (\cite{inoue00}, IHK) have
derived a formula for the conversion from dust IR luminosity to
SFR by using the following three quantities: the fraction of
Lyman continuum luminosity absorbed by gas ($f$), the fraction
of UV luminosity absorbed by dust ($\epsilon$), and the
fraction of dust heating from old ($\ga 10^8$ yr) stellar
populations ($\eta$). We develop a method to estimate those
three quantities based on the idea that the various way of SFR
estimates from ultraviolet (UV) luminosity (2000 \AA\
luminosity), H$\alpha$ luminosity, and dust IR luminosity
should return the same SFR. After applying our method to
samples of galaxies, the following results are obtained in our
framework. First, our method is applied to a sample of
star-forming galaxies, finding that $f\sim 0.6$,
$\epsilon\sim 0.5$, and $\eta\sim 0.4$ as representative
values. Next, we apply the method to a starburst sample, which
shows larger extinction than the star-forming galaxy sample.
With the aid of $f$, $\epsilon$, and $\eta$, we are able to
estimate reliable SFRs from UV and/or IR luminosities.
Moreover, the H$\alpha$ luminosity, if the H$\alpha$ extinction
is corrected by using the Balmer decrement, is suitable for a
statistical analysis of SFR, because the same correction
factor for the Lyman continuum extinction (i.e., $1/f$) is
applicable to both normal and starburst galaxies over all the
range of SFR.
The metallicity dependence of $f$ and $\epsilon$ is also tested:
Only the latter proves to have
a correlation with metallicity.  As an extension of
our result, the local ($z=0$) comoving density of SFR can be
estimated with our dust extinction corrections. We show that all
UV, H$\alpha$, and IR comoving luminosity densities at $z=0$ give
a  consistent SFR per comoving volume
($\sim 3~10^{-2}h~\solyr~{\rm Mpc}^{-3}$). Useful formulae for SFR
estimate are listed.
 }
\titlerunning{Extinction and star formation rate}
\authorrunning{H. Hirashita et al.}
\maketitle
%
%________________________________________________________________

%% the body of the paper

\section{Introduction}\label{sec:intro}

During the history of the universe, galaxies have evolved forming
stars. As a result, the present universe is filled with a large
amount of stars and a significant amount of radiative energy
originating from such stars. Therefore, tracing the star formation
activity over all the history in the universe is fundamental to
understand how the present universe has formed. To quantify the
star formation activity,  Star Formation Rate (SFR), defined as
the stellar mass formed per unit time, is often estimated.

The star formation activity, more specifically the SFR, can be
traced by young
($\mbox{age}\la 10^{7}$--$10^8$ yr)\footnote{In this paper,
the word ``young'' is used to specify the timescale on which the
current SFR is traced. The word ``old'' is used otherwise.} stars.
Since short-lived massive stars are certain to be produced most
recently, the SFR is traced with the luminosity of massive stars.
One of the well known and commonly used tracers of massive stars
is H$\alpha$ luminosity (e.g., Kennicutt \cite{kennicutt83}),
because H$\alpha$ photons originate from the gas ionised by
massive-star radiation. Since massive stars are the strong source
for Ultra-Violet (UV) photons, UV luminosity is also used as an
indicator of SFR. A
general review for SFR indicators can be found in
Kennicutt (\cite{kennicutt98a}).

In order to obtain a reliable estimate of the SFR, we have
to take into account dust absorption
(extinction)\footnote{Strictly speaking, the extinction is
defined as the sum of scattering and absorption. In this paper,
the term ``dust extinction'' is used to indicate
only the absorption.} of UV and H$\alpha$ light. The H$\alpha$
luminosity (or more generally,
the luminosity of hydrogen recombination lines) is decreased also
by the extinction of Lyman continuum photons (e.g.,
Smith et al.\ \cite{smith78};
Inoue et al.\ \cite{inouehk01}; Inoue \cite{inoue01};
Charlot \& Fall \cite{charlot00};
Charlot \& Longhetti \cite{charlot01};
Charlot et al.\ \cite{charlot02};
Dopita et al.\ \cite{dopita03}). Quantifying the dust
absorption is generally crucial in deriving the star formation
history of a galaxy (e.g.,
Inoue et al.\ \cite{inoue01}; Hopkins et al.\ \cite{hopkins01};
Kewley et al.\ \cite{kewley02}) or of the universe (e.g.,
Flores et al.\ \cite{flores99};
Steidel et al.\ \cite{steidel99}; Meurer et al.\ \cite{meurer99}).
Nevertheless, the correction for dust absorption usually
requires an elaborate multi-wavelength or multi-band modelling
(e.g., Calzetti \cite{calzetti01}).

There are some SFR indicators that are virtually free from dust
extinction. One of them is the infrared (IR) luminosity
originating from dust continuum emission
(Kennicutt \cite{kennicutt98a}). In this paper, we use the term IR
to indicate the wavelength range where
dust emission dominates the luminosity ($\sim 8$--1000 $\mu$m). We
call the total luminosity of dust emission ``dust IR luminosity.''
Contrary to UV and H$\alpha$, dust IR luminosity traces the stellar
radiation absorbed by dust. Thus, if a significant fraction of
stellar radiation is absorbed by dust, detecting the dust IR
emission is important. Recent {\it ISO}\, (e.g.,
Takeuchi et al.\ \cite{takeuchi01}), {\it COBE}\, (e.g.,
Hauser \& Dwek \cite{hauser01}), and SCUBA observations
(e.g., Blain et al.\ \cite{blain99};
Barger et al.\ \cite{barger00}) have shown a significant
contribution of dust IR emission to the total light in the
universe over a large part of the cosmic history.
%%The radio emission originating from supernova remnants can
%%be used as an extinction-free SFR indicator (e.g.,
%%Condon \cite{condon92}). We do not use the radio luminosity in
%%this paper, but in the future it could be included
%%as a natural extension of our method described in this paper.

Theoretically the conversion factor between the dust IR luminosity
and SFR is dependent on how efficiently stellar light is absorbed
by dust and reprocessed in IR (e.g.,
Inoue et al.\ \cite{inoue00}, hereafter IHK). In the analytic
formula of IHK, the conversion factor between IR luminosity and
SFR are described by using the following three
parameters: $f$, $\epsilon$, and $\eta$, where $f$ is the fraction
of Lyman continuum luminosity absorbed by hydrogen atoms,
$\epsilon$ is the nonionising photons from young stars
absorbed by dust, and $\eta$ is the fraction of IR luminosity
originating from dust heating by old stars.

Let us note some remarks about the three parameters. IHK assume
that the ionizing photons do not escape out of galaxies; that is,
the fraction $1-f$ of the Lyman continuum
luminosity is absorbed by dust grains. This can be justified for
nearby galaxies, whose escape fraction of Lyman
continuum photons is generally less than 10\% (e.g.,
Leitherer et al.\ \cite{leitherer95};
Deharveng et al.\ \cite{deharveng01}). For high-redshift (high-$z$)
galaxies, it is still a matter of debate whether the escape fraction
is high (Steidel et al.\ \cite{steidel01}) or low
(Heckman et al.\ \cite{heckman01};
Giallongo et al.\ \cite{giallongo02};
Fern\'{a}ndez-Soto et al.\ \cite{fernandez03}). 
We can regard
$\epsilon$ as the fraction of UV (912--3650 \AA)
photons absorbed by dust,
partly because almost all the radiative energy from young
stars lies in the UV range, and partly because the dust
absorption is much more efficient in UV than in optical
(Buat \& Xu \cite{buat96}).
With respect to $\eta$, we should keep in mind that the definition
of $\eta$ is temporal, not spatial as ``cirrus''. Indeed, young
stars can heat dust outside H\,{\sc ii} regions and it is shown
that the UV
heating is also efficient in all the interstellar medium
(ISM) via a diffuse interstellar radiation field not spatially
associated with H\,{\sc ii} regions (Buat \& Xu \cite{buat96};
Walterbos \& Greenawalt \cite{walterbos96}). Our definition of
$\eta$ is also different from the cool dust fraction as defined
in Lonsdale Persson \& Helou (\cite{lonsdale87}), because dust
located out of H\,{\sc ii} regions can be heated by UV radiation
from massive stars and at the same time it can remain cool.
Because of such a temporal definition of $\eta$, we have to
specify the timescale on which the current SFR is traced
(Section \ref{subsec:reconsideration}).

It is known that IR/UV flux ratio is a good indicator for the
dust absorption in UV (Buat et al.\ \cite{buat99};
Meurer et al.\ \cite{meurer99}; Witt \& Gordon \cite{witt00};
Panuzzo et al.\ \cite{panuzzo03}). Therefore, UV and IR
luminosities are useful not only to derive SFR but also to correct
UV luminosity for dust absorption. By using
H$\alpha$ luminosity in addition to UV and IR luminosities, we
develop a
method to estimate $f$, $\epsilon$, and $\eta$ (the principal
quantities in the IHK formalism), and propose a way to obtain
a reliable estimate of the SFR.
Rosa-Gonz\'{a}lez et al.\ (\cite{rosa02}) also
treat $\epsilon$ to estimate the SFR from dust IR luminosity,
but we stress that $f$ and $\eta$ are also important (IHK).
Charlot et al.\ (\cite{charlot02}) have compared the UV, IR,
and H$\alpha$ SFR estimators based on a fully consistent model,
which includes different extinctions between young and
old stellar populations, non-Balmer emission lines, and various
types of star formation histories (see also Charlot \& Fall
\cite{charlot00}). Charlot et al.\  (\cite{charlot02}) also use
[O {\sc ii}] line luminosity (see also
Gallagher et al.\ \cite{gallagher89}), which is not included in
our paper, to estimate SFR This paper adopts a different
approach: instead of modelling the details of radiative
processes, we
develop an independent and simple way to extract the important
quantities for SFR estimates, so that our model might be
applied easily to large data sets.

Hirashita et al.\ (\cite{hiro01}, hereafter H01) suggest that those
three quantities, $f$, $\epsilon$, and $\eta$, change as galaxies
are enriched by metals. As a galaxy forms stars and recycles gas
into ISM, the metallicity increases in some classes of models
such as a closed-box model (e.g., Tinsley \cite{tinsley_sola80}).
At the same time, dust grains are made from metals. In fact,
metallicity is related to dust content (e.g.,
Issa et al.\ \cite{issa90};
Schmidt \& Boller \cite{schmidt93};
Lisenfeld \& Ferrara \cite{lisenfeld98}; Dwek \cite{dwek98};
Hirashita \cite{hiro99}; Edmunds \cite{edmunds01}). H01 consider
that the metallicity evolution results in the increase
of dust optical depth. Consequently $f$ and $\epsilon$ (and
possibly $\eta$) can be affected by metallicity evolution.

The aim of this paper is to develop a method to observationally
estimate the quantities important for determining SFR. The
luminosities in this paper are derived by assuming an isotropic
radiation, which we expect to be reasonable in a statistical
sense. First we reconstruct the IHK formula to make it consistent
with our treatment in this paper
(Section \ref{sec:reconstruction}). In
Section \ref{sec:method}, then, we explain how to derive the
important quantities ($f$, $\epsilon$, and $\eta$) for estimating
SFR. In Section \ref{sec:sample}, we present the samples to which
we apply our method. The statistical properties of various SFR
indicators are discussed in Section \ref{sec:result}, in which
the metallicity dependence of $f$ and $\epsilon$ is also tested.
In Section \ref{sec:summary}, we summarise our results and
discuss possible application of our method to the cosmic Star
Formation History (SFH) and to
future survey data. 

\section{Reconstruction of SFR formulae}\label{sec:reconstruction}

\subsection{Reconsideration of the formula between $L_{\rm IR}$ and
SFR}\label{subsec:reconsideration}

We reconstruct the conversion formula from dust IR luminosity
($L_{\rm IR}$) to SFR following the method of IHK. [In
Dale et al.\ (\cite{dale01}), this luminosity is called ``Total IR
luminosity (TIR).''] A subset of the \iras\ sample is
commonly used to investigate the IR properties of galaxies, but
the \iras\ data are limited to the wavelength shorter than
120 $\mu$m. Therefore, the conversion formula from \iras\
luminosity defined between 40 and 120 $\mu$m to the total dust
IR luminosity is useful. We call the luminosity in the
40--120 $\mu$m range ``FIR luminosity'' ($L_{\rm FIR}$), whose
way of estimate can be seen in
Lonsdale Persson \& Helou (\cite{lonsdale87}).
Recently, data at longer wavelengths by {\it ISO}\, have enabled
Dale et al.\ (\cite{dale01}) to estimate the ratio
$L_{\rm IR}/L_{\rm FIR}$ as a function of the \iras\
60 $\mu$m vs.\ 100 $\mu$m flux ratio. Hereafter we will use this
their result unless otherwise stated. The ratio
$L_{\rm IR}/L_{\rm FIR}$ is larger ($>2$) for normal star-forming
galaxies (such as spiral galaxies) than for
starburst galaxies. While the past analyses based on starburst
models (e.g., Meurer et al.\ \cite{meurer99}) are not affected by
a small difference between $L_{\rm IR}$ and $L_{\rm FIR}$
($L_{\rm IR}/L_{\rm FIR}\simeq 1.4$; but see
Calzetti et al.\ \cite{calzetti00}, who derive
$L_{\rm IR}/L_{\rm FIR}\simeq 1.8$)\footnote{For our {\it IUE}
sample, which we treat as a starburst sample, the application of
Dale et al.\ (\cite{dale01}) to our data leads to
$L_{\rm IR}/L_{\rm FIR}\simeq 2.0$ on average. The uncertainty
in $L_{\rm IR}$ by 50\% causes the change of $f$, $\epsilon$,
and $\eta$ by $\sim 20$\%.}, we have to be careful about the
difference for normal galaxies.

The current star formation activity of galaxies can be traced
by quantifying the amount of young stars.
Thus, the term ``young'' should indicate the timescale on
which the current star formation is traced. This timescale is
denoted as $t_{\rm SF}$. For example,
Inoue (\cite{inoue02a}) adopts $t_{\rm SF}\sim 10^7$ yr to trace
the SFR. It is important to adopt a common
timescale for all the SFR tracers. Since our aim is to work with
monochromatic data near 2000 \AA, the best choice of the
timescale is that appropriate for this wavelength. We choose
$t_{\rm SF}=10^8$ yr unless otherwise stated because the
2000 \AA\ luminosity reaches its stationary value around $10^8$ yr
in a constant SFR. Moreover,
for large galaxies active in star formation but not
necessarily starbursting, a constant SFR over  $10^8$ yr seems
reasonable since the strong
correlation found between the H$\alpha$ and UV emissions (e.g.,
Buat et al.\ \cite{buat02}, hereafter B02) argue for such a
stationarity (but the stationarity for other samples is not
necessarily supported; Sullivan et al.\ \cite{sullivan01}). Such a
hypothesis strongly simplifies the analysis and helps the
understanding. Since the duration of the current star
formation activity may be shorter in starburst galaxies, we also
examine $t_{\rm SF}=10^7$ yr in
Section \ref{subsec:uncertainty}.

IHK have established a procedure to derive the conversion
formula which connects dust IR luminosity and SFR. They start
from the following relation among luminosities originating from
young stars (Petrosian et al.\ \cite{petrosian72}):
\begin{eqnarray}
L_{\rm IR}^{\rm SF}=L_{{\rm Ly}\alpha}+(1-f)L_{\rm Lyc}+\epsilon
L_{\rm nonion}\, ,\label{eq:petrosian}
\end{eqnarray}
where $L_{\rm IR}^{\rm SF}$, $L_{{\rm Ly}\alpha}$, $L_{\rm Lyc}$,
and $L_{\rm nonion}$ are various kinds of luminosities originating
from young stars
(luminosities of dust IR, Ly$\alpha$, Lyman continuum, and
nonionising photons, respectively), and $f$ and $\epsilon$ are
the fraction of Lyman continuum luminosity absorbed by gas
and the fraction of nonionising photon luminosity absorbed
by dust, respectively. In this formalism, all the Ly$\alpha$
photons are assumed to be absorbed by dust grains during some
resonant scatterings. This assumption can be justified even for
dust-deficient galaxies
with 1\% of the Galactic dust-to-gas ratio (H01), but we should
be careful about this point if we apply our formula to objects
from which
a large amount of Ly$\alpha$ photons leak for some reason.
As we mentioned in Section \ref{sec:intro}, it is also assumed
that the Lyman continuum photons are
absorbed either by gas or by dust and do not escape out of
the galaxy.

All the three luminosities on the right-hand side in equation
(\ref{eq:petrosian}) are related to the bolometric luminosity of
the young stars, $L_{\rm bol}^{\rm SF}$. In order to obtain such
relations, we have run Starburst99
(Leitherer et al.\ \cite{leitherer99}) and made synthetic stellar
spectra. We adopt the Salpeter stellar initial mass
function (IMF) with the upper and lower masses of 100 $M_\odot$ and
0.1 $M_\odot$, respectively,
constant SFR, and solar metallicity. We use the result at the
age of $10^8$ yr, and this timescale should be equal to
$t_{\rm SF}$. Then, we finally obtain
$L_{\rm Lyc}=0.13L_{\rm bol}^{\rm SF}$ and
$L_{\rm nonion}=0.87L_{\rm bol}^{\rm SF}$ from the synthetic
spectrum at $t_{\rm SF}=10^8$ yr.
In this paper, we adopt the same parameter set for the
Starburst99 spectrum unless otherwise stated.

The Ly$\alpha$ luminosity is estimated under Case B
(Osterbrock \cite{osterbrock89}) as (IHK)
\begin{eqnarray}
L_{\rm Ly\alpha}=\frac{2}{3}N_{\rm Lyc}f'h\nu_{\rm Ly\alpha}\, ,
\end{eqnarray}
where $N_{\rm Lyc}$ is the number of ionising photons emitted
per unit time, $h\nu_{\rm Ly\alpha}$ is the energy of a
Ly$\alpha$ photon ($1.63\times 10^{-11}$ erg), and $f'$ is the
{\it number fraction} of Lyman continuum photons absorbed by gas
(note that $f$ is the {\it luminosity
fraction} of Lyman continuum absorbed by gas). Although the
relation between $f$ and $f'$ depends in a minor way on the
extinction law for Lyman continuum photons and the
shape of the Lyman continuum spectrum, $f\simeq f'$ can be
expected. There is a large difficulty in modelling the
relation between $f$ and $f'$ because of the lack of
knowledge about the extinction law for the Lyman continuum
photons. Thus, we simply adopt $f=f'$ throughout this paper. By
using the conversion from $N_{\rm Lyc}$ to $L_{\rm Lyc}$
predicted by the synthesised spectrum of the Starburst99
($N_{\rm Lyc}=3.22\times 10^{10}L_{\rm Lyc}$ in the cgs units),
we obtain $L_{\rm Ly\alpha}=0.34fL_{\rm Lyc}$. Then,
equation (\ref{eq:petrosian}) is reduced to
\begin{eqnarray}
L_{\rm IR}^{\rm SF}=(0.13-0.085f+0.87\epsilon )
L_{\rm bol}^{\rm SF}\, .\label{eq:SF_bol}
\end{eqnarray}

The Starburst99 result indicates that
\begin{eqnarray}
\frac{\rm SFR}{\solyr}=1.79\times 10^{-10}\,
\frac{L_{\rm bol}^{\rm SF}}{L_\odot}\, .\label{eq:SFR_bol}
\end{eqnarray}
The dust IR luminosity, $L_{\rm IR}$, is the sum of two components
originating from young ($<t_{\rm SF}$) stars and old
($>t_{\rm SF}$) stars (remember that $t_{\rm SF}=10^8$ yr unless
otherwise stated). Then the fraction of old stellar contribution
to $L_{\rm IR}$, $\eta$, is defined as
\begin{eqnarray}
L_{\rm IR}^{\rm SF}=(1-\eta )L_{\rm IR}\, .\label{eq:def_cirrus}
\end{eqnarray}
Considering the energy balance of dust, $\eta$ is also interpreted
as the fraction of the energy input into dust (i.e., dust heating)
from the old stellar population. Using
equations (\ref{eq:SF_bol}), (\ref{eq:SFR_bol}), and
(\ref{eq:def_cirrus}), we obtain
\begin{eqnarray}
\frac{\rm SFR}{\solyr}=
\frac{1.79\times 10^{-10}(1-\eta )}{0.13-0.085f+0.87\epsilon}\,
\frac{L_{\rm IR}}{L_\odot}\, .\label{eq:conversion_IR}
\end{eqnarray}
For the following convenience, we express the above conversion
formula as
\begin{eqnarray}
{\rm SFR}=C_{\rm IR}(f,\,\epsilon ,\,\eta )\, L_{\rm IR}\, ,
\label{eq:conversion_formula}
\end{eqnarray}
that is, the conversion factor $C_{\rm IR}$ becomes
\begin{eqnarray}
C_{\rm IR}(f,\,\epsilon ,\,\eta )=
\frac{1.79\times 10^{-10}(1-\eta )}
{0.13-0.085f+0.87\epsilon}~[\solyr~L_\odot^{-1} ]\, .
\label{eq:def_cir}
\end{eqnarray}
We observe that the conversion factor is determined by a set
of $(f,\,\epsilon ,\,\eta )$. We should note that this
numerical expression for the conversion factor is based on the
assumption that the star formation occurs at a constant rate
for $10^8$ yr.

\subsection{SFR from IR and UV emissions}

We formally define SFR(IR) as
\begin{eqnarray}
{\rm SFR(IR)}=C_{\rm IR}^{\rm sb}\,
L_{\rm IR}\, ,\label{eq:SFRIR}
\end{eqnarray}
where we define the conversion factor, $C_{\rm IR}^{\rm sb}$,
as follows:
\begin{eqnarray}
C_{\rm IR}^{\rm sb} & \equiv & C_{\rm IR}
(f=0,\,\epsilon =1,\,\eta =0)\nonumber\\
& = & 1.79\times 10^{-10}~\solyr~L_\odot^{-1}\, .
\end{eqnarray}
If the radiation field in a galaxy is dominated by the young
stars and all the radiation from stars is once absorbed by
dust and reemitted in IR, SFR(IR) gives a reliable
estimate for SFR (IHK). Indeed,
$L_{\rm IR}=L_{\rm bol}^{\rm SF}$ in such a
case, and equation (\ref{eq:SFRIR}) is the same as
equation (\ref{eq:SFR_bol}). This situation may be realised in
starburst galaxies as assumed in Kennicutt (\cite{kennicutt98b}).
Thus, we call the condition
$(f,\,\epsilon ,\,\eta)=(0,\, 1,\, 0)$
``dusty starburst approximation.'' $C_{\rm IR}^{\rm sb}$
strongly depends on $t_{\rm SF}$ (Appendix \ref{app:sfr1e7}).

The intrinsic UV luminosity can be related to the SFR in a rather
straightforward way. We adopt the 2000 \AA\ monochromatic
luminosity $L_{2000}$ to trace UV (B02), and we express the SFR
formula as
\begin{eqnarray}
{\rm SFR(UV)}=C_{\rm 2000}\, L_{\rm 2000}\, ,\label{eq:SFRUV}
\end{eqnarray}
where $C_{2000}$ can be calculated by using the Starburst99
spectrum (with the same parameters as above in Section
\ref{subsec:reconsideration}) without dust absorption:
$C_{2000}=2.03\times 10^{-40}~(\solyr )/$(erg s$^{-1}$ \AA$^{-1}$).
The advantage of using 2000 \AA\ monochromatic luminosity is that
(i) 2000 \AA\ is roughly the centre of the UV wavelength range
(i.e., it traces the mean property of
UV luminosity and extinction) and (ii) a large number of UV
data are available at 2000 \AA. If there is no dust
absorption, SFR(UV) gives a reliable estimate of SFR.

Both SFR(IR) and SFR(UV) have their own disadvantages. If
we know $\eta$, $(1-\eta)\,{\rm SFR(IR)}$ can be
estimated from the observed $L_{\rm IR}$. However,
$(1-\eta)\,{\rm SFR(IR)}$ would systematically underestimate
the SFR, because a part of the
stellar radiation is not absorbed by dust. This underestimate
can be supplemented by SFR(UV) because most of the unabsorbed
light from young stars is in UV. On the other hand,
SFR(UV) can be formally estimated by multiplying the observed
2000 \AA\ monochromatic luminosity with $C_{2000}$, but this
always underestimates the SFR because an appreciable amount of the
radiation from young stars is absorbed by dust and reprocessed into
IR. This underestimate can be supplemented by
$(1-\eta)\,{\rm SFR(IR)}$. Therefore, the following sum
of UV and IR SFRs is expected to give a better
approximation of the SFR:
\begin{eqnarray}
{\rm SFR}\simeq (1-\eta )\,{\rm SFR(IR)}+{\rm SFR(UV)}\, .
\label{eq:SFR_IR_UV}
\end{eqnarray}
In the previous works such as Flores et al.\ (\cite{flores99})
and Buat et al.\ (\cite{buat99}), a simple sum
${\rm SFR(IR)}+{\rm SFR(UV)}$ has been adopted. This
type of simple sum
is examined in Section \ref{subsec:sfr_comp}.

The formula for SFR(IR) (eq.\ \ref{eq:SFRIR}) is derived by
assuming that all the stellar light is absorbed by dust. On the
other hand, the conversion to SFR(UV) (eq.\ \ref{eq:SFRUV}) is
made by assuming a theoretical stellar spectrum without
extinction. Thus, we expect that equation (\ref{eq:SFR_IR_UV})
gives a
reasonable estimate for the SFR if there is no dust absorption
or if the dust optical depth against the stellar light is
significantly larger than 1. In the case where there is some dust
absorption, the shape of the absorbed spectrum in UV is modified
by the differential
absorption, and the IR light comes from only a part of the UV
radiative energy. Therefore, it is not obvious whether or not
equation (\ref{eq:SFR_IR_UV}) is valid for an arbitrary value
of dust extinction (e.g., $E(B-V)$).

\begin{figure}
\begin{center}
\includegraphics[width=8cm]{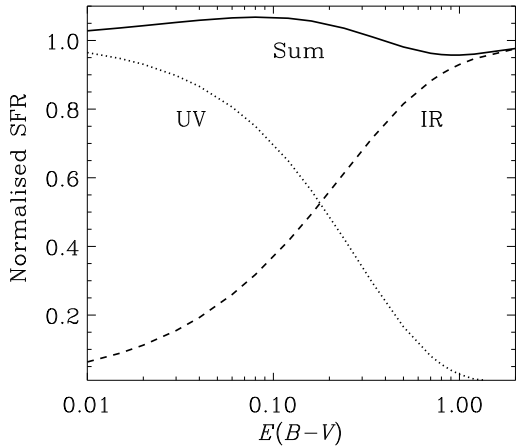}
\end{center}
\caption{Star formation rates (SFRs) normalised with the true
star formation rate. We show the part of the SFRs traced with UV
and IR luminosities (dotted and dashed lines, respectively).
The Calzetti extinction curve is assumed. The solid line shows
the sum of the two SFRs and the correct SFR (normalised) should
be 1. The small deviation of the solid line
from 1 means that the sum of UV and IR SFRs traces the true SFR.}
\label{fig:sfr_sum}
\end{figure}

We check the validity of equation (\ref{eq:SFR_IR_UV}).
We start from the Starburst99 spectrum. Here, we only consider
the contribution from young stars (i.e., $\eta =0$), but
we can apply the following consideration to $\eta >0$
by replacing SFR(IR) and $L_{\rm IR}$ with
$(1-\eta )\,{\rm SFR(IR)}$ and $(1-\eta )\, L_{\rm IR}$,
respectively. Assuming the Calzetti extinction curve
(Calzetti et al.\ \cite{calzetti00}) for the dust absorption, we
extinguish the synthesised stellar spectrum as a function of
$E(B-V)$.
The energy absorbed by dust is assumed to be
equal to $L_{\rm IR}$. The energy absorbed by dust is estimated
over all the range of the Starburst99 spectrum
(100--1,600,000 \AA)\footnote{Almost all
($\sim 90$\%) of the absorbed energy originates from the UV
range.} by using
Calzetti et al.'s fitting formula. As a result, we obtain
$L_{2000}$ (reduced according to the extinction) and $L_{\rm IR}$,
and then SFR(IR) and SFR(UV) from
equations (\ref{eq:SFRIR}) and (\ref{eq:SFRUV}), respectively. In
this modelling, both SFR(IR) and SFR(UV) are proportional to SFR
given in the Starburst99 calculation. Thus, we should examine if
${\rm [SFR(IR)+SFR(UV)]/SFR}=1$ is well satisfied.
In Figure \ref{fig:sfr_sum}, we show SFR(IR)/SFR, SFR(UV)/SFR,
and ${\rm [SFR(IR)+SFR(UV)]/SFR}$ as a function of $E(B-V)$
(dashed, dotted, and solid lines, respectively). The deviation of
${\rm [SFR(IR)+SFR(UV)]/SFR}$ from 1 is very small ($\la 6$\%).

We have used the Calzetti curve because it is the only extinction
curve derived from a sample of galaxies. But its validity is
checked only for starburst galaxies. However, even if we adopt
the Milky Way extinction curve
(Cardelli et al.\ \cite{cardelli89}; with $R_V=3.1$)
instead of the Calzetti one, the difference of
${\rm [SFR(IR)+SFR(UV)]/SFR}$ from 1 is also $\la 7$\%. Thus,
we expect that the selection of a specific extinction curve does
not change our conclusion that the sum of the IR and UV SFRs is
an excellent indicator for the SFR.

It may be worth mentioning qualitatively the following details,
whose quantitative discussion nevertheless depends on the
assumed stellar spectrum and extinction curve. The slight
overestimate of SFR for $E(B-V)\la 0.1$ is due to a smaller
extinction (i.e., a larger escape fraction) of the 2000 \AA\
light than that of the total UV light. Because of this, the
2000 \AA\ monochromatic luminosity slightly overestimates the
UV SFR. This point
can be seen later in Figure \ref{fig:epscomp}
(Section \ref{subsec:epsilon}). On the contrary, SFR(IR) tends to
underestimate the SFR even in the high-extinction limit as we can
see for $E(B-V)\ga 0.5$. This is because a small part of energy
escapes from galaxies in optical and near-infrared wavelengths.

\section{Estimates of the principal parameters}\label{sec:method}

\begin{figure*}
\begin{center}
\includegraphics[width=15cm]{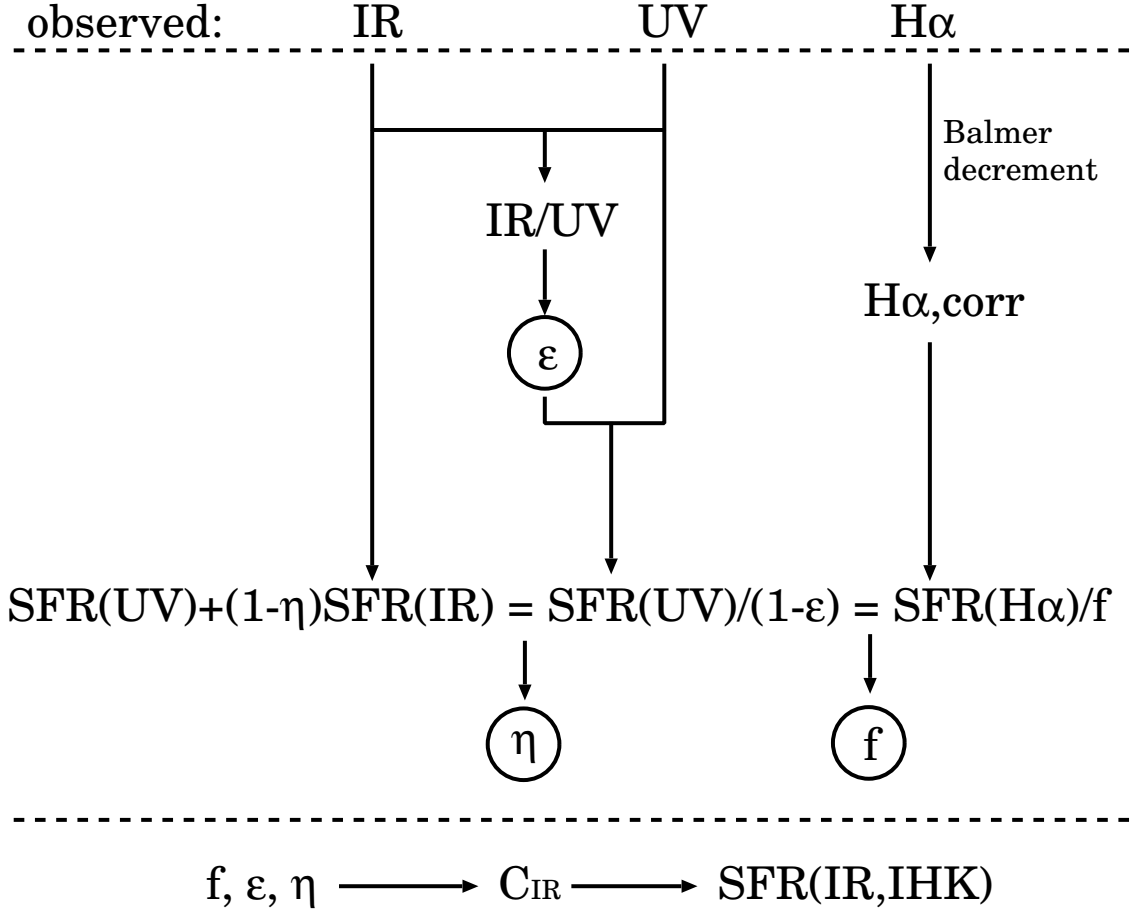}
\end{center}
\caption{Summary of our method. We use three observational
data (dust IR, UV, and H$\alpha$ luminosities) and derive
three parameters $f$, $\epsilon$, and $\eta$, which
are important for the conversion factor ($C_{\rm IR}$) from dust
IR luminosity to star formation rate (SFR).
The method is based on the consistency of the various SFRs
derived from (1) the combination of UV and dust IR luminosities,
(2) UV luminosity corrected for dust extinction, and
(3) H$\alpha$ luminosity corrected for dust extinction.
Finally, using the three parameters, we derive $C_{\rm IR}$,
which gives another independent estimate for SFR (denoted as
SFR(IR,\,IHK)).}
\label{fig:chart}
\end{figure*}

We explain how to determine the parameters $f$, $\epsilon$,
and $\eta$, all of which appear in the IHK formula 
(eq.\ \ref{eq:conversion_IR}). We use the UV (2000 \AA),
H$\alpha$, and dust IR luminosities to derive those three
quantities. We illustrate our procedures in
Figure \ref{fig:chart}.

\subsection{$\epsilon$}\label{subsec:epsilon}

The parameter $\epsilon$ is the fraction of UV photons absorbed
by dust. Here, the UV wavelength range is defined between
912 \AA\ and 3650 \AA. The intrinsic UV luminosity is denoted
as $L_{\rm UV}^0$.
The observational UV luminosity $L_{\rm UV}$ can then be
expressed as
\begin{eqnarray}
L_{\rm UV}=(1-\epsilon )\, L_{\rm UV}^0\, .\label{eq:observedUV}
\end{eqnarray}
Both $L_{\rm UV}$ and $L_{\rm UV}^0$ are defined as the
luminosities in the UV wavelength range.

As mentioned in Section \ref{sec:reconstruction}, we use the
2000 \AA\ monochromatic luminosity to estimate the UV SFR.
Therefore, we have to relate $\epsilon$ to $\epsilon_{2000}$
(defined as the fraction of 2000 \AA\ monochromatic luminosity
absorbed by dust). This relation depends on assumed models.
One of the simplest ways to avoid such model dependence is to
assume $\epsilon_{2000}=\epsilon$.

\begin{figure}
\begin{center}
\includegraphics[width=8cm]{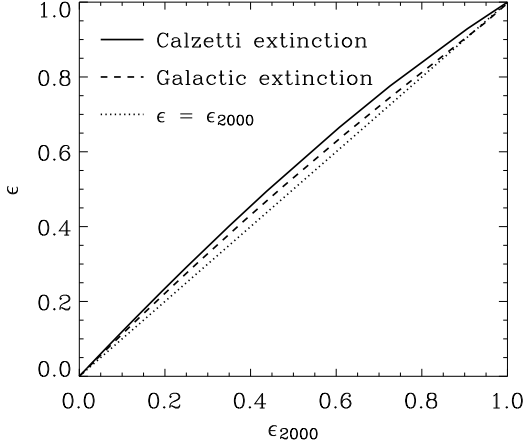}
\end{center}
\caption{Relation between $\epsilon$ (fraction of UV luminosity
absorbed by dust) and $\epsilon_{2000}$ (fraction of 2000 \AA\
monochromatic luminosity absorbed by dust) modelled for two
types of extinction curves. The solid and dashed lines show the
result for the
Calzetti and Galactic extinction curves. The dotted line
presents $\epsilon =\epsilon_{2000}$.}
\label{fig:epscomp}
\end{figure}

We quantify the uncertainty of this assumption
$\epsilon_{2000}=\epsilon$. We adopt the Calzetti extinction
curve to determine the absorbed fraction of UV light as a
function of the extinction at 2000 \AA. For a certain value
of the 2000 \AA\ extinction and the Starburst99 spectrum, we
obtain the monochromatic luminosity after dust absorption for
each wavelength. By integrating the monochromatic luminosity
in all the UV wavelength range (912 -- 3650 \AA) and comparing
this integration with the integration of the spectrum before dust
absorption, we obtain the fraction of UV light absorbed by dust.
Consequently, we obtain $\epsilon$ as a function of
$\epsilon_{2000}$. The
$\epsilon$ -- $\epsilon_{2000}$ relation is plotted in
Figure \ref{fig:epscomp} (solid line). We find that
$\epsilon_{2000}$ tends to underestimate $\epsilon$ but that the
difference between the two is at most $\sim 0.06$. The difference
in $\epsilon$ propagates to the estimates of $f$ and $\eta$ in our
framework described later, causing the uncertainty of at most
$\sim 20$\% in those two parameters. However, we should keep in
mind that the conversion from $\epsilon$ to $\epsilon_{2000}$
always has uncertainty if a specific model
(i.e., a set of extinction curve and intrinsic spectrum) is
selected. For example, if the Galactic extinction curve
(Cardelli et al.\ 1989; with $R_V=3.1$) is assumed, we obtain
the dashed line in Figure \ref{fig:epscomp}. In this case also,
the difference from $\epsilon =\epsilon_{2000}$ is small
enough. We prefer to avoid the model dependence by simply
adopting $\epsilon_{2000}=\epsilon$.

The IR/UV flux ratio can be used to estimate $\epsilon$
(Meurer et al.\ \cite{meurer99}; Buat et al.\ \cite{buat99}; B02).
In principle, we can derive the extinction at 2000 \AA\
from a reasonable model of the radiative transfer such as
Xu \& Buat (\cite{xu95}) if all the required data are available.
Since the treatment of radiative transfer is beyond the scope
of our simple analysis, we adopt a simplified relation
between IR/UV ratio and 2000 \AA\ extinction.
For the SFG sample (see Section \ref{sec:sample} for the sample
description), we use the calibration of Buat et al.\ (\cite{buat99})
but expressed in
$F_{\rm IR}/F_{2000}$ (dust IR vs.\ 2000 \AA) flux ratio instead of
$F_{\rm FIR}/F_{2000}$ (\iras\ FIR vs.\ 2000 \AA) one since the
extrapolation of the total IR flux from that observed by \iras\
has been made possible with the ISOPHOT observations (Dale et al.\
\cite{dale01}). We adopt the same definition of $F_\lambda$
($\lambda$ is a wavelength in UV in units of \AA) as that in
Buat et al.\ (\cite{buat99}), i.e,
$F_\lambda\equiv\lambda f_\lambda$, where
$f_\lambda$ is the flux density per wavelength. Then
we finally derive
\begin{eqnarray}
A_{2000} & \simeq & 0.21+0.75\, \log(F_{\rm IR}/F_{2000})\nonumber\\
         & + & 0.47[\log(F_{\rm IR}/F_{2000})]^2\, ,
\end{eqnarray}
where $A_{2000}$ (mag) is the extinction at 2000 \AA. We note that 
$F_{\rm IR}/F_{2000}=L_{\rm IR}/(\lambda_{2000}\, L_{\rm 2000})$,
where $\lambda_{2000}=2000$ \AA.
This relation is derived from the energetics between IR and UV
and is found rather insensitive to the details of the stellar
spectrum if there is such an ongoing star formation activity
as is seen in our samples
(Buat \& Xu \cite{buat96}). The fraction of 2000 \AA\ light 
absorbed by dust can be related to $A_{2000}$ as
\begin{eqnarray}
\epsilon =\epsilon_{2000} = 1-10^{-A_{2000}/2.5}\, .
\label{eq:eps_a2000}
\end{eqnarray}

We also investigate starburst galaxies observed by {\it IUE}\,
as described in Section \ref{sec:sample}.
For this {\it IUE} sample, we will use the
extinction estimated by 
Calzetti et al.\ (\cite{calzetti00}). 
The extinction at 1600 \AA\ can be observationally estimated
for the {\it IUE}\, sample as (Calzetti \cite{calzetti01})
\begin{eqnarray}
A_{1600}\simeq 2.5\log\left[\frac{1}{1.68}\,
\frac{F_{\rm IR}}{F_{1600}}+1\right]\, ,
\end{eqnarray}
where $F_{1600}$ is defined in the same way as $F_{2000}$, i.e.,
the flux density at 1600 \AA\ multiplied by 1600 \AA.
For nearby galaxies, $F_{1600}\simeq F_{2000}$
(Deharveng et al.\ \cite{deharveng94}; Buat et al.\ \cite{buat99}),
and thus we can use IR vs.\ 2000 \AA\ luminosity ratio to estimate
$F_{\rm IR}/F_{1600}$. Since $A_{2000}=0.9A_{1600}$
under the Calzetti extinction curve, $\epsilon_{2000}$, which is
assumed to be equal to $\epsilon$, is estimated by
\begin{eqnarray}
\epsilon =\epsilon_{2000}=1-10^{-0.9A_{1600}/2.5}\, .
\end{eqnarray}

\subsection{$\eta$}\label{subsec:eta}

We start from equation (\ref{eq:SFR_IR_UV}), where the basic idea
is that the SFR measured with the IR emission is the SFR 
lost from the UV light because of the extinction. The left-hand
side of equation (\ref{eq:SFR_IR_UV}) can be estimated by
correcting the UV luminosity for dust absorption, i.e.,
\begin{eqnarray}
{\rm SFR}={\rm SFR(UV)}/(1-\epsilon )\, .\label{eq:SFR_corr_UV}
\end{eqnarray}
For the right-hand side of equation (\ref{eq:SFR_IR_UV}), SFR(IR)
is estimated from $L_{\rm IR}$ (eq.\ \ref{eq:SFRIR}) while SFR(UV)
from $L_{2000}$ (eq.\ \ref{eq:SFRUV}). Since
$\epsilon$ is known after applying the method in
Section \ref{subsec:epsilon}, $\eta$ is determined from the
following equation derived from equations (\ref{eq:SFRIR}),
(\ref{eq:SFRUV}), (\ref{eq:SFR_IR_UV}) and (\ref{eq:SFR_corr_UV}):
\begin{eqnarray}
(1-\eta)\, C_{\rm IR}^{\rm sb}\, L_{\rm IR}=
\frac{\epsilon}{1-\epsilon}\, C_{2000}\, L_{2000}\, .
\end{eqnarray}

\subsection{$f$}

We start from the following relation between SFR and Lyman
continuum luminosity:
\begin{eqnarray}
{\rm SFR}=C_{L_{\rm Lyc}}\, L_{\rm Lyc}\, ,
\end{eqnarray}
where the Starburst99 spectrum indicates that
$C_{L_{\rm Lyc}}=3.45\times 10^{-43}~(\solyr )/({\rm erg~s}^{-1})$.
We also obtain
\begin{eqnarray}
{\rm SFR}=C_{N_{\rm Lyc}}\, N_{\rm Lyc}\, ,\label{eq:SFR_Nc}
\end{eqnarray}
where $N_{\rm Lyc}$ is the number of the ionising photons
emitted per unit time, and
$C_{N_{\rm Lyc}}=1.07\times 10^{-53}~(\solyr )/{\rm s}^{-1}$
according to the Starburst99 spectrum.

Since the H$\alpha$ luminosity traces the amount of ionising
photons, we connect the above expressions to the
H$\alpha$ luminosity.
According to Deharveng et al.\ (\cite{deharveng01}),
\begin{eqnarray}
fN_{\rm Lyc}=7.34\times 10^{11}L_{\rm H\alpha}^{\rm c}\, ,
\label{eq:caseb}
\end{eqnarray}
where 
$L_{\rm H\alpha}^{\rm c}$ is the H$\alpha$ luminosity
corrected for dust absorption by using the Balmer decrement
(i.e., observed H$\alpha$ luminosity is multiplied by
$10^{A({\rm H}\alpha )/2.5}$, where
$A({\rm H}\alpha)$ (mag) is the extinction for the H$\alpha$
photons), and the
quantities are expressed in the cgs units. This relation is
determined from the Case B condition.
Then we finally obtain the
following relation from
equations (\ref{eq:SFR_Nc}) and (\ref{eq:caseb}):
\begin{eqnarray}
{\rm SFR}=C_{\rm H\alpha}\, L_{\rm H\alpha}^{\rm c}/f\, ,
\label{eq:formula_ha}
\end{eqnarray}
where we obtain numerically
$C_{\rm H\alpha}=7.34~10^{11}\,C_{N_{\rm Lyc}}=7.89~10^{-42}$
$(\solyr )/({\rm erg~s}^{-1})$. The conversion
factors $C_{L_{\rm Lyc}}$, $C_{N_{\rm Lyc}}$, and
$C_{\rm H\alpha}$ are similar to the past estimates (e.g., 
Kennicutt \cite{kennicutt98a}).

We estimate SFR in the left-hand side in
equation (\ref{eq:formula_ha}) from UV luminosity, dust IR
luminosity, and $\eta$ (known by the method in
Section \ref{subsec:eta}) by using
equation (\ref{eq:SFR_IR_UV}). By using
equations (\ref{eq:SFRIR}) and (\ref{eq:SFRUV}) in addition, $f$
is obtained:
\begin{eqnarray}
f=
\frac{C_{\rm H\alpha}\, L_{\rm H\alpha}^{\rm c}}
{(1-\eta )\, C_{\rm IR}^{\rm sb}\,
L_{\rm IR}+C_{2000}\, L_{2000}}\, .
\label{eq:f}
\end{eqnarray}

\section{Sample selection}\label{sec:sample}

One of the samples of nearby star-forming galaxies whose
extinction properties and star formation rates are well
examined is
the SFG sample of B02. The sample consists of spiral and
irregular galaxies located in clusters (Coma, Abell 1367,
Cancer and Virgo). B02 also treat a starburst sample observed by
{\it IUE}. We also use this sample ({\it IUE}\, sample) in order
to test the applicability of our method and to investigate the
difference in properties. We adopt only the galaxies with a good
measurement of Balmer decrement (i.e., with a direct measurement
of the underlying stellar absorption).
For the {\it IUE} sample, we only use the galaxies whose
angular diameter is less than 1.5 arcmin in order to avoid the
small aperture problem of {\it IUE}. The original quantities and
the details are listed in Tables 1 and 2 of B02 for the SFG
sample and the {\it IUE}\, sample, respectively (see also the
references therein).

The observational quantities used for our analysis are
$L_{\rm H\alpha}^{\rm c}$ (the H$\alpha$ luminosity corrected for
dust absorption by using the Balmer decrement), $L_{2000}$
(the monochromatic luminosity at 2000 \AA), and $L_{\rm IR}$
(the dust IR luminosity). The dust IR luminosity is converted
from the \iras\
FIR luminosity by the method described in
Dale et al.\ (\cite{dale01}). The three quantities for the SFG
sample and the {\it IUE} sample are listed
in Tables \ref{tab:sfg_sample} and \ref{tab:sb_sample},
respectively. We also examine the dependence on metallicity, which
is traced with the oxygen abundance. The oxygen abundance is taken
from Gavazzi et al.\ (2003, in preparation) for the SFG sample
and from Calzetti et al.\ (\cite{calzetti94}) for the {\it IUE}
sample. We call $12+\log{\rm (O/H)}$ metallicity in this paper.
The solar metallicity corresponds to $12+\log{\rm (O/H)}=8.93$
(Anders \& Grevesse \cite{anders89}; Cox \cite{cox00}).

\begin{table*}
\begin{center}
\caption{SFG sample. The sample is taken from Buat et al.\ (2002).}
\begin{tabular}{p{0.2\linewidth}cccccc}\hline\noalign{\smallskip}
Name & $L_{\rm 2000}$ & $L_{\rm H\alpha}^{\rm c}$ & $L_{\rm IR}$ &
$f$ & $\epsilon$ & $\eta$ \\
 & erg s$^{-1}$ \AA$^{-1}$ & erg s$^{-1}$ & erg s$^{-1}$ & & & \\
 \noalign{\smallskip}\hline\noalign{\smallskip}
VCC 25     & 8.91e+39 & 2.45e+41 & 6.76e+43 & 0.51 & 0.52 &    0.36  \\
VCC 66     & 2.34e+39 & 1.12e+41 & 1.95e+43 & 0.84 & 0.54 &    0.36  \\
VCC 89     & 7.94e+39 & 2.14e+41 & 7.08e+43 &    0.46 & 0.56 & 0.36  \\
VCC 92     & 5.37e+39 & 2.69e+41 & 5.01e+43 &    0.83 & 0.57 & 0.37  \\
VCC 131    & 1.02e+39 & 6.46e+39 & 4.17e+42 &    0.16 & 0.36 & 0.39  \\
VCC 307    & 8.13e+39 & 7.24e+41 & 2.00e+44 &    0.80 & 0.77 & 0.41  \\
VCC 318    & 2.04e+39 & 3.39e+40 & 4.79e+42 &    0.50 & 0.22 & 0.48  \\
VCC 459    & 6.31e+38 & 1.41e+40 & 1.17e+42 &    0.73 & 0.16 & 0.55  \\
VCC 664    & 8.51e+38 & 1.74e+40 & 2.24e+42 &    0.60 & 0.25 & 0.45  \\
VCC 692    & 7.94e+38 & 1.32e+40 & 4.37e+42 &    0.36 & 0.44 & 0.37  \\
VCC 801    & 2.40e+39 & 1.05e+41 & 3.02e+43 &    0.61 & 0.64 & 0.38  \\
VCC 827    & 1.15e+39 & 3.31e+40 & 2.63e+43 &    0.27 & 0.76 & 0.40  \\
VCC 836    & 1.29e+39 & 1.20e+41 & 3.89e+43 &    0.72 & 0.80 & 0.41  \\
VCC 938    & 9.12e+38 & 3.16e+40 & 6.03e+42 &    0.69 & 0.49 & 0.36  \\
VCC 1189   & 6.31e+38 & 1.00e+40 & 1.62e+42 &    0.47 & 0.24 & 0.46  \\
VCC 1205   & 1.66e+39 & 2.19e+40 & 8.71e+42 &    0.29 & 0.43 & 0.37  \\
VCC 1379   & 1.82e+39 & 2.57e+40 & 8.13e+42 &    0.34 & 0.39 & 0.38  \\
VCC 1450   & 1.55e+39 & 2.40e+40 & 7.08e+42 &    0.36 & 0.39 & 0.38  \\
VCC 1554   & 4.07e+39 & 2.24e+41 & 3.47e+43 &    0.96 & 0.55 & 0.36  \\
VCC 1678   & 6.92e+38 & 1.29e+40 & 1.12e+42 &    0.63 & 0.13 & 0.61  \\
CGCG 97087  & 3.72e+40 & 9.33e+41 & 2.14e+44 & 0.53 & 0.45 & 0.37  \\
CGCG 100004 & 3.80e+39 & 1.55e+41 & 3.39e+43 & 0.69 & 0.56 & 0.36  \\
CGCG 119029 & 3.16e+39 & 2.29e+41 & 5.75e+43 & 0.80 & 0.71 & 0.39  \\
CGCG 119041 & 5.62e+38 & 8.13e+40 & 5.37e+43 & 0.43 & 0.92 & 0.44  \\
CGCG 119043 & 1.48e+39 & 1.00e+41 & 3.63e+43 & 0.61 & 0.77 & 0.41  \\
CGCG 119046 & 4.79e+39 & 2.19e+41 & 2.95e+43 & 0.94 & 0.47 & 0.36  \\
CGCG 119047 & 3.55e+39 & 1.95e+41 & 7.76e+43 & 0.54 & 0.75 & 0.40  \\
CGCG 119054$^{\rm a}$ & 1.86e+39 & 2.34e+41 & 2.04e+43 &    1.90 &
          0.61 &   0.37 \\
CGCG 119059 & 1.23e+39 & 4.37e+40 & 2.75e+43 & 0.34 & 0.75 & 0.40  \\
CGCG 160055 & 1.26e+40 & 3.39e+41 & 1.91e+44 & 0.34 & 0.68 & 0.38  \\
CGCG 160067 & 5.62e+39 & 3.09e+41 & 5.89e+43 & 0.85 & 0.60 & 0.37  \\
CGCG 160139 & 1.02e+40 & 2.29e+41 & 4.17e+43 & 0.55 & 0.36 & 0.39  \\
CGCG 160252 & 5.01e+39 & 3.80e+41 & 1.86e+44 & 0.50 & 0.83 & 0.42  \\
\noalign{\smallskip}
\hline\noalign{\smallskip}
Mean       &          &          &          &    0.57 & 0.53 & 0.40  \\
$\sigma$ & & &  & 0.21 & 0.21 & 0.06 \\ \noalign{\smallskip}
\hline
\end{tabular}
\begin{list}{}{}
\item[$^{\mathrm{a}}$] This galaxy is not considered in taking the
mean and $\sigma$ because $f$ is larger than 1.
\end{list}
\label{tab:sfg_sample}
\end{center}
\end{table*}

\begin{table*}
\begin{center}
\caption{{\it IUE} starburst sample. The sample is compiled in
Table 2 of Buat et al.\ (2002). We only list and use galaxies whose
angular size is less than 1.5 arcmin in order to avoid the small
aperture effect.}
\begin{tabular}{p{0.2\linewidth}cccccc}\hline\noalign{\smallskip}
Name & $L_{\rm 2000}$ & $L_{\rm H\alpha}^{\rm c}$ & $L_{\rm IR}$ &
$f$ & $\epsilon$ & $\eta$ \\
 & erg s$^{-1}$ \AA$^{-1}$ & erg s$^{-1}$ & erg s$^{-1}$ & & &  \\
 \noalign{\smallskip}\hline\noalign{\smallskip}
Mrk 499 & 8.51e+39 & 4.57e+41 & 2.00e+44 &  0.32 &  0.85 & $-$0.03 \\
Mrk 357  & 6.46e+40 & 2.09e+42 & 3.89e+44 &  0.50 &  0.60 & $-$0.12 \\
IC 1586$^{\rm a}$ & 4.37e+39 & 7.76e+41 & 8.71e+43 & 1.20 & 0.83 & $-$0.05
       \\
Mrk 66   & 4.90e+39 & 1.02e+41 & 4.37e+43 &  0.25 &  0.69 & $-$0.10   \\
NGC 5860$^{\rm a}$ & 3.47e+39 & 1.05e+42 & 1.26e+44 & 1.27 & 0.89 & $-$0.00
     \\
UGC 9560 & 6.17e+38 & 2.69e+40 & 2.82e+42 & 0.78 & 0.54 & $-$0.13 \\
NGC 6090 & 1.20e+40 & 3.80e+42 & 8.71e+44 & 0.74 & 0.94 & 0.05 \\
IC 214  & 1.02e+40 & 1.15e+42 & 1.05e+45 &  0.20 &  0.96 &  0.08 \\
Tol 1924$-$416 & 6.76e+39 & 2.75e+41 & 2.51e+43 & 0.81 & 0.49 & $-$0.14 \\
Haro 15   & 1.38e+40 & 3.02e+41 & 1.23e+44 &  0.26 &  0.69 & $-$0.10 \\
NGC 6052 & 5.50e+39 & 3.89e+41 & 2.75e+44 &  0.23 &  0.92 &  0.02 \\
NGC 3125 & 5.62e+38 & 3.89e+40 & 7.59e+42 &  0.63 &  0.77 & $-$0.07 \\
NGC 1510 & 2.34e+38 & 5.50e+39 & 1.23e+42 &  0.39 &  0.57 & $-$0.12 \\
NGC 1614 & 3.98e+39 & 5.62e+42 & 1.78e+45 &  0.67 &  0.99 &  0.20 \\
NGC 7673 & 5.01e+39 & 4.37e+41 & 1.17e+44 &  0.52 &  0.85 & $-$0.03 \\
NGC 7250 & 8.71e+38 & 2.88e+40 & 1.07e+43 &  0.32 &  0.75 & $-$0.08 \\
NGC 5996 & 1.38e+39 & 1.74e+41 & 5.37e+43 &  0.50 &  0.90 &  0.00  \\
NGC 1140 & 1.70e+39 & 9.33e+40 & 1.62e+43 &  0.63 &  0.70 & $-$0.09 \\
NGC 4194$^{\rm a}$ & 2.40e+39 & 1.91e+42 & 3.39e+44 & 1.04 & 0.97 & 0.11 \\
\noalign{\smallskip}
\hline\noalign{\smallskip}
Mean       &          &          &          &  0.48 & 0.76 & $-0.04$ \\
$\sigma$ & & & & 0.20 & 0.15 & 0.09 \\ \noalign{\smallskip}
\hline
\end{tabular}
\begin{list}{}{}
\item[$^{\mathrm{a}}$] Those galaxies are not considered in taking the
mean and $\sigma$ because $f$ is larger than 1.
\end{list}
\label{tab:sb_sample}
\end{center}
\end{table*}

{}From the arguments on dust temperature and equivalent widths
of the Balmer lines, B02 show that the {\it IUE}\, sample
consists of galaxies with higher star formation activity than the
SFG sample. The {\it IUE}\, sample can be regarded as a class of
``starburst'' galaxies, while the SFG sample can be
representative of ``normal'' star-forming galaxies (spiral
galaxies and irregular galaxies). We use the term ``normal''
and ``starburst'' to indicate a rough classification of
star formation activity in this paper, and those two classes are
often identified with the SFG sample and the
{\it IUE}\, sample, respectively.

\section{Results}\label{sec:result}

In this section, we first apply our method to the SFG sample
(Section \ref{subsec:result_SFG}), because the {\it IUE}\, sample
may still have an effect of the small aperture even after we put
the criterion for the angular size. The robustness of the method
against the model is also examined for the SFG sample
(Section \ref{subsec:uncertainty}). Then, we apply our method
to the {\it IUE}\, sample in Section \ref{subsec:result_IUE}.
The SFRs derived from the IHK method is compared with the
best-estimate SFR in order to test the validity of the IHK
formula (Section \ref{subsec:validIHK}). Other SFR estimators
are also
examined in Section \ref{subsec:sfr_comp}. The metallicity
dependence of some of the quantities ($f$ and $\epsilon$) is
investigated in Section \ref{subsec:metallicity} in order to test
the hypothesis in H01. The luminosities and the SFR conversion
factors are summarised in Table \ref{tab:notation}.

\begin{table*}
\begin{center}
\caption{Luminosities and SFR conversion factors on the
timescale of $t_{\rm SF}=10^8$ yr.}
\begin{tabular}{p{0.15\linewidth}ccl}\hline\noalign{\smallskip}
Quantity & Value & Units & Definition / comment \\
 \noalign{\smallskip}\hline\noalign{\smallskip}
$L_{2000}$ & & erg s$^{-1}$ \AA$^{-1}$ & 2000 \AA monochromatic \\
$L_{\rm H\alpha}^{\rm c}$ & & erg s$^{-1}$ & corrected for Balmer
decrement \\
$L_{\rm IR}$ & & erg s$^{-1}$ & dust IR$^{\rm a}$ \\
\noalign{\smallskip}
\hline\noalign{\smallskip}
$C_{2000}$ & $2.03\times 10^{-40}$ $^{\rm b}$ & 
$(\solyr)$/(erg s$^{-1}$ \AA$^{-1}$) & to be divided by
 $(1-\epsilon )$ \\
$C_{\rm H\alpha}$ & $7.89\times 10^{-42}$ $^{\rm c}$ &
$(\solyr)$/(erg s$^{-1}$) &
 to be divided by $f$ \\
$C_{\rm IR}^{\rm sb}$ & $1.79\times 10^{-10}$ $^{\rm d}$ &
$\solyr~L_\odot^{-1}$ & dusty starburst approximation for IR \\
$C_{\rm IR}$ & equation (\ref{eq:def_cir}) & $\solyr~L_\odot^{-1}$ &
 $f$, $\epsilon$, and $\eta$ are necessary \\
\noalign{\smallskip}
\hline
\end{tabular}
\begin{list}{}{}
\item[$^{\mathrm{a}}$] The total luminosity of dust emission
derived from the \iras\ luminosity correcting for
the contribution from longer ($\lambda >120~\mu$m) wavelengths.
Dale et al.\ (\cite{dale01}) is used for the correction,
but the difference between correction models is well less than
30\% (e.g., the difference from Nagata et al.\ \cite{nagata02}).
\item[$^{\mathrm{b}}$] If $t_{\rm SF}=10^7$ yr,
the value becomes $3.18\times 10^{-40}$.
\item[$^{\mathrm{c}}$] The value is unchanged for
$t_{\rm SF}=10^7$ yr.
\item[$^{\mathrm{d}}$] If $t_{\rm SF}=10^7$ yr,
the value becomes $2.72\times 10^{-10}$.
\end{list}
\label{tab:notation}
\end{center}
\end{table*}

\subsection{$f$, $\epsilon$, and $\eta$ for the SFG sample}
\label{subsec:result_SFG}

One of the characteristics of the SFG sample is that the
H$\alpha$ (corrected for dust absorption by the Balmer decrement) to UV
flux ratio is lower than that expected for $f=1$ under a constant
SFR over $10^8$ yr (B02). This may indicate that some significant
fraction of ionising photons is absorbed by dust grains.
Inoue (\cite{inoue01}), after analysing the individual H\,{\sc ii}
regions of some Local Group galaxies, also reaches the same
conclusion. Charlot et al.\ (\cite{charlot02}) also
concluded the same thing from their analysis of the
Stromlo-APM redshift survey data.
Thus, we expect to find a fraction $f$ significantly
lower than 1.

In Table \ref{tab:sfg_sample}, we list $f$, $\epsilon$, and $\eta$
for each galaxy. By definition, $f$ should be between 0 and 1, but
only CGCG 119054 shows $f$ significantly larger than 1. We
consider this to be due to the overcorrection of H$\alpha$
absorption. Indeed, $A({\rm H}\alpha )=2.47$ mag is the largest
of all the SFG sample and the correction of dust absorption is as
large as a factor of 9.7. With such a large extinction, the
extinguished H$\beta$ flux measurement could be very uncertain. As
a result,
the extinction derived from the Balmer decrement
measurement could have a significant uncertainty. Therefore, we
omit CGCG 119054 in the following analysis.

The mean values and the standard deviations ($\sigma$) of $f$,
$\epsilon$, and $\eta$ are shown at the bottom of
Table \ref{tab:sfg_sample}. We calculated those values excluding
CGCG 119054. As expected at the beginning of this subsection, $f$
is significantly smaller than 1. The mean value of $f$ (0.57)
indicates that about 40\% of the ionising photons are directly
absorbed by dust grains before being processed into
recombination lines. This can be a reason for the systematic
underestimate of H$\alpha$ SFR relative to SFRs from other
indicators (e.g., Cram et al.\ \cite{cram98}; B02). We also
observe from the mean of $\epsilon$ (0.53) that the half of the UV
is absorbed by dust and reprocessed into IR. If we convert
$\epsilon =0.53\pm 0.21$ into $A_{2000}$ by using
equation (\ref{eq:eps_a2000}) and assuming
$\epsilon =\epsilon_{2000}$ (i.e., by
$A_{2000}=-2.5\log (1-\epsilon )$), we obtain
$A_{2000}=0.82^{+0.64}_{-0.40}$. This confirms the result of
B02. The value of $\eta\sim 0.4$ indicates that about 40\% of the
dust heating is due to old (age $>t_{\rm SF}=10^8$ yr) stars,
which have nothing to do with the current SFR.
Misiriotis et al.\ (\cite{misiriotis01}) also find
the contribution of old stellar populations to grain
heating to be $\sim 40$\% for a sample of spiral galaxies.

\begin{figure*}
\begin{center}
\includegraphics[width=8cm]{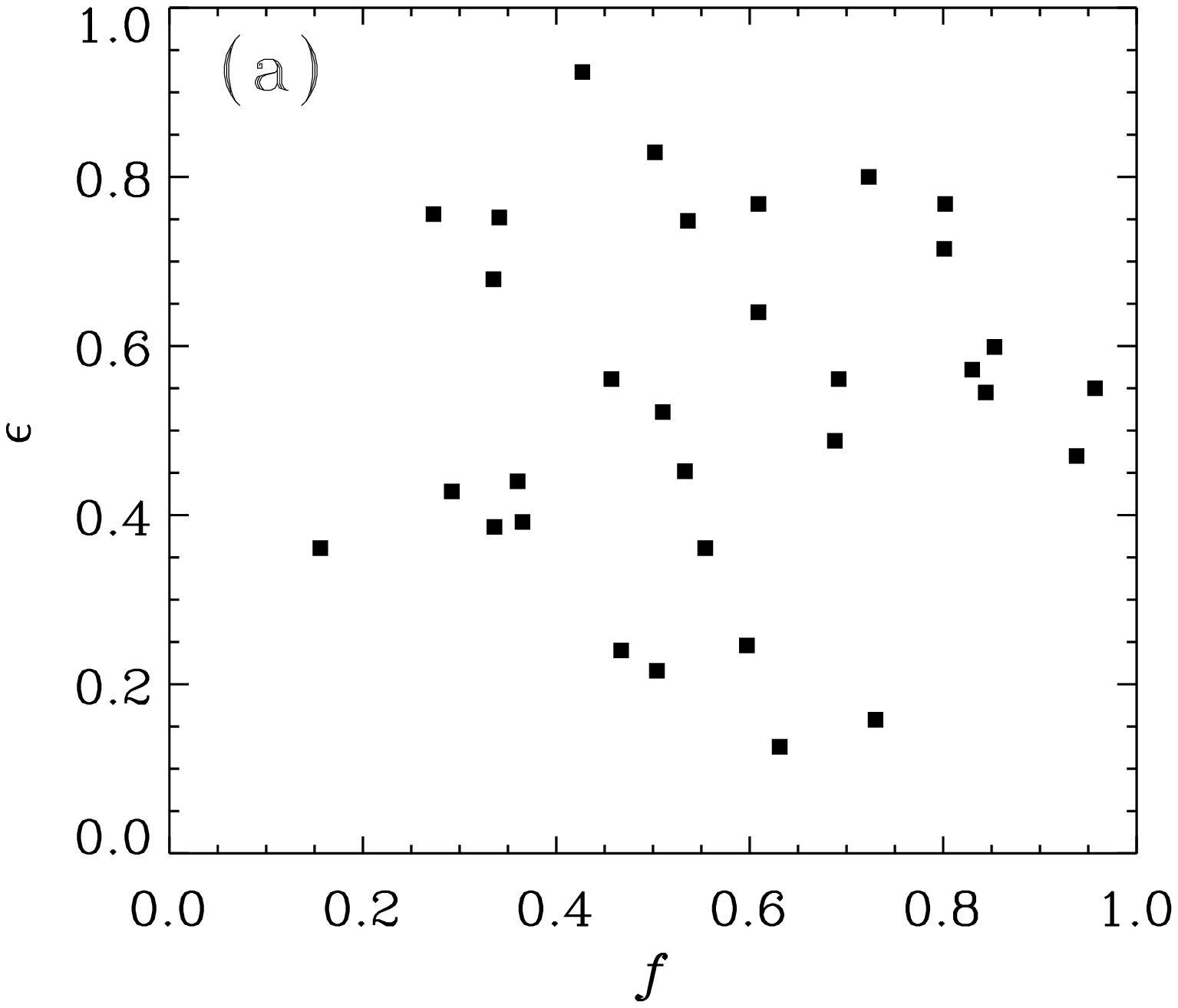}
\includegraphics[width=8cm]{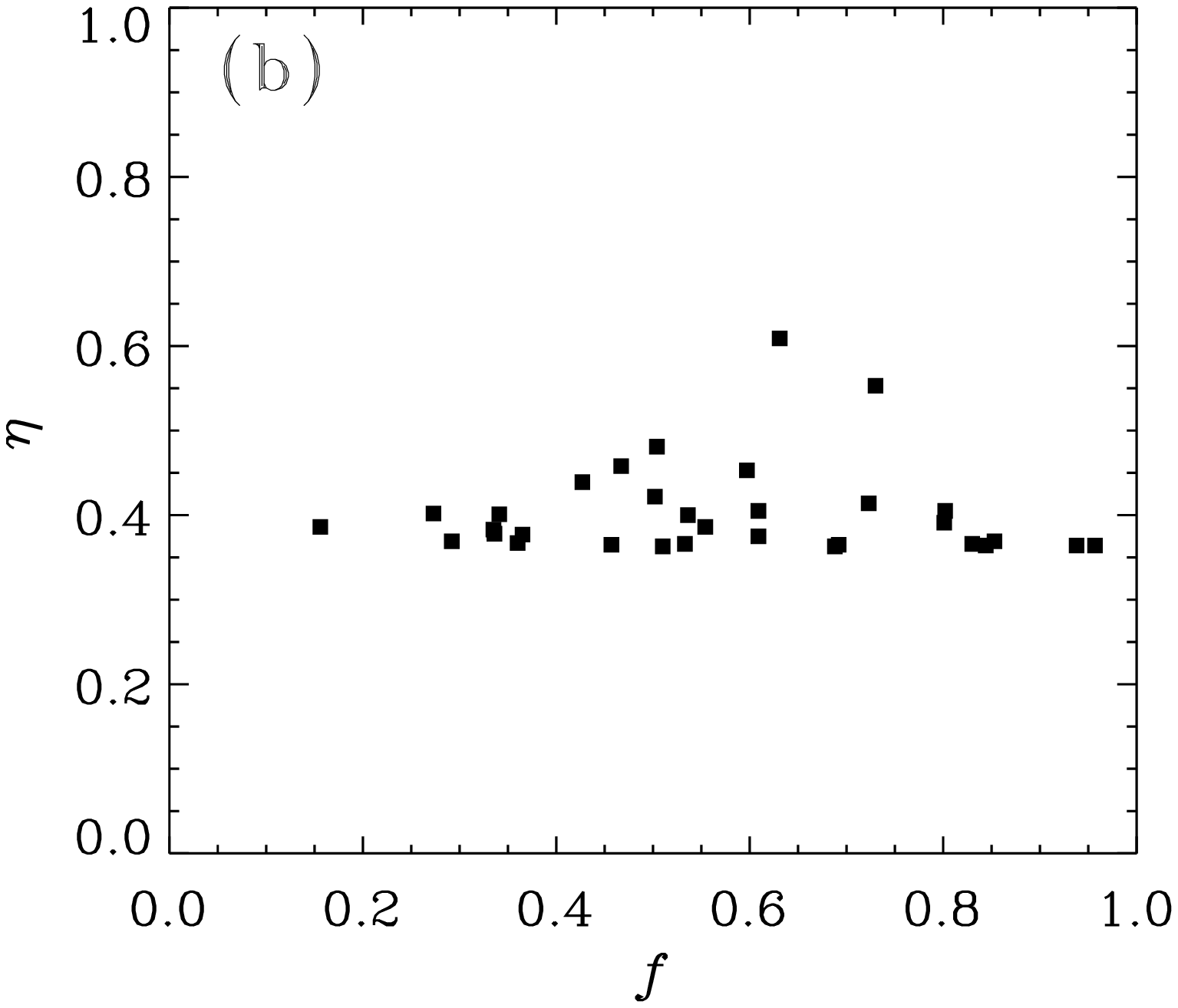}\\
\includegraphics[width=8cm]{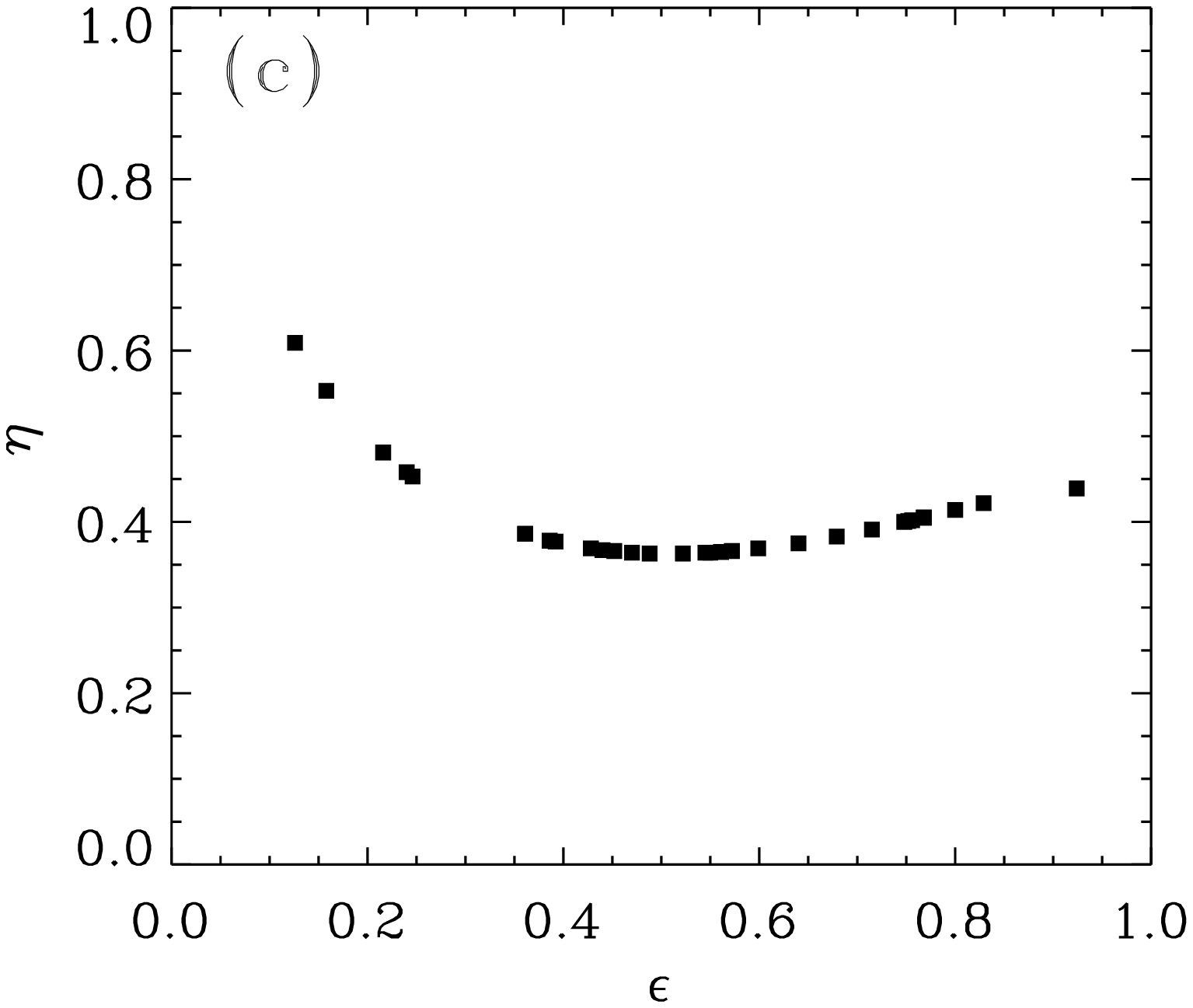}
\end{center}
\caption{Diagrams of (a) $f$ vs.\ $\epsilon$,
(b) $f$ vs.\ $\eta$, and (c) $\eta$ vs.\ $\epsilon$ for the SFG
sample, where $f$ is the fraction of ionising radiation absorbed
by gas, $\epsilon$ is the fraction of UV luminosity absorbed by
dust, and $\eta$ is the fraction of dust heating by the old
($>10^8$ yr) stellar population.}
\label{fig:correlation}
\end{figure*}

We also show the relations among the three quantities in
Figure \ref{fig:correlation}. We see that there is no evidence
for correlation either between $f$ and $\epsilon$ ($r=0.055$;
$r$ is the correlation coefficient) or between $f$ and $\eta$
($r=-0.017$). There seems to be a tight relation between $\eta$
and $\epsilon$, but this tightness results from our
formulation that determines both $\eta$ and $\epsilon$
as a function of IR/UV flux ratio (see
Sections \ref{subsec:epsilon} and \ref{subsec:eta}). In reality,
there is a scatter in the relation between $\epsilon$ and IR/UV
flux ratio (see Fig.\ 1 of Buat et al.\ \cite{buat99}). This
scatter also disperse the $\eta$--$\epsilon$ relation and the
standard deviation of $\eta$ would increase significantly.
However, the mean values of $\eta$ should still be $\sim 0.4$.

\subsection{Model uncertainty}\label{subsec:uncertainty}

We examine how much the three quantities ($f$, $\epsilon$, and
$\eta$) are changed by the uncertainty in model assumptions.
One of the sources for the uncertainty is the conversion
between the \iras\ FIR luminosity to the (total) dust IR
luminosity. We have used
Dale et al.\ (\cite{dale01}) for this conversion. If we adopt
Nagata et al.\ (\cite{dale02}) for this conversion (their
FIR2 is used), the mean values of $f$, $\epsilon$, and $\eta$
change to 0.62, 0.50, and 0.39, respectively (the standard
deviations being 0.24, 0.20, and 0.06, respectively).
The difference in $L_{\rm IR}/L_{\rm FIR}$ between
Dale \& Helou (\cite{dale02}) and Dale et al.\ (\cite{dale02}) is
smaller than that between Nagata et al.\ (\cite{nagata02})
and Dale et al.\ (\cite{dale01}). Therefore, with the available
$L_{\rm IR}/L_{\rm FIR}$ models, the parameters are determined
within an uncertainty of $\sim 10$\%.

Rosa-Gonz\'{a}lez et al.\ (\cite{rosa02}) also calculated
$\epsilon$. Some of their galaxies (13 galaxies) overlap with our
{\it IUE}\, sample, whose $\epsilon$ is derived later
(Section \ref{subsec:result_IUE}; Table \ref{tab:sb_sample}). They
determined $\epsilon$ in a similar manner as ours, but not the
same (e.g., the way of the estimate of IR/UV flux ratio is
different). The agreement between our $\epsilon$
and theirs is extremely good (the difference is within
0.05 except for Mrk 66, for which we obtain $\epsilon =0.69$
while they derive $\epsilon =0.58$). This supports the
robustness of $\epsilon$ for the {\it IUE}\, sample.
For the SFG sample, the literature that
analysed $\epsilon$ is not found, and we cannot argue that
other models support our derivation of $\epsilon$.

We should remember that the conversion factors between SFR and
various luminosities depend on age, or more generally on SFH
(e.g., Sullivan et al.\ \cite{sullivan01}). In particular, the
relative luminosity ratio between UV and Lyman continuum (or
recombination lines) is sensitive to age. In a constant SFR, the
luminosity of the Lyman continuum reaches its stationary value at
the age of $\sim 10^7$ yr, while the UV luminosity becomes
stationary at $\sim 10^8$ yr. In the above, we have adopted
$t_{\rm SF}=10^8$ yr for the age, but we could adopt
$t_{\rm SF}=10^7$ yr to have an idea how much the quantities
change in response to the SFH.
Some of the conversion factors at $10^7$ yr change:
$C_{2000}$, $C_{\rm IR}$, and $C_{\rm IR}^{\rm sb}$ take the values
described in Appendix \ref{app:sfr1e7}, while $C_{L_{\rm Lyc}}$,
$C_{N_{\rm Lyc}}$, and $C_{\rm H\alpha}$ are the same as those at
$10^8$ yr.

In the framework of Starburst99,
it is very difficult to have a spectrum for an arbitrary SFH,
but it is easy and computationally economical to change only the
age under a constant SFH. Thus, we only change the age
here. By using another population synthesis code, we have also
tested the difference between
a constant SFR and an exponentially decaying SFR. As long as the
exponential decaying timescale is comparable to (or longer than) the
duration of the star formation ($t_{\rm SF}$), the change of th
results is not so drastic as the difference between
$t_{\rm SF}=10^8$ yr and $t_{\rm SF}=10^7$ yr. If a galaxy has a
decaying timescale much shorter than $t_{\rm SF}$, it would not
be classified with a star-forming galaxy and would not included
in our sample.

If those coefficients for $t_{\rm SF}=10^7$ yr are adopted, we
obtain the mean values $f=0.39$, $\epsilon =0.53$, and
$\eta =0.38$ (the standard deviations, $\sigma$, being 0.20, 0.21,
and 0.06, respectively). It is natural that $\epsilon$ does not
change at all because it is determined from IR/UV ratio, which
is independent of the conversion factors. While $\eta$ is
not sensitive to the conversion factors, $f$ changes
significantly depending on the assumed $t_{\rm SF}$.
Recalling that $f$ is determined from the ratio between the
SFR traced with H$\alpha$ and the SFR traced with UV and IR,
the sensitive change of $f$ against $t_{\rm SF}$ comes from
the difference in the conversion factors for the UV and IR SFRs.
Therefore, the typical age of the current star-forming activity
is important if the age is shorter than $10^8$ yr. We expect
that our SFG sample has a continuous mode of
the star formation because the correlation between H$\alpha$
and UV luminosities is good (B02). However, for a starburst
sample such as the {\it IUE}\, sample, the duration of the
present starburst could be shorter than $10^8$ yr.
Calzetti et al.\ (\cite{calzetti94}) assumed a constant SFR
over 2 $10^7$ yr for these galaxies.
The correlation between H$\alpha$ and UV luminosities is not so
good as that for the SFG galaxies (B02), which also suggests
that the sample has a diverse property in SFH on a timescale
shorter than $10^8$ yr. However, since we do not know
the typical age of those sample, we adopt $t_{\rm SF}=10^8$ yr
also for this sample. If $t_{\rm SF}$ is shorter than
$10^8$ yr, $f$ becomes smaller. Some mechanisms are
proposed for a short-term variation of SFR on a galactic
scale (Kamaya \& Takeuchi \cite{kamaya97} and references
therein).

The spectral synthesis model used to derive the conversion
factors is another source of uncertainty. We have seen above
that the relative values between $C_{\rm H\alpha}$ and
$C_{2000}$ (or $C_{\rm IR}$) is the largest source for the
uncertainty in $f$, although $\eta$ is determined quite
robustly. Based on {\sc pegase} synthesis code
(Fioc \& Rocca-Volmerange \cite{fioc97}), for example, the
construction of the conversion factors are possible. Two of the
conversion factors ($C_{\rm H\alpha}$ and $C_{2000}$ in our
notation) are thus obtained by
Sullivan et al.\ (\cite{sullivan01}) for various age and
metallicity. However, both of their $C_{2000}$ and
$C_{\rm H\alpha}$ are similar to ours (difference is $<20$\%).
Therefore, the difference in $f$ between the two synthesis
codes is small and the difference in $t_{\rm SF}$ causes larger
difference in $f$. The difference in metallicity
affects $C_{\rm H\alpha}$ as largely as that in $t_{\rm SF}$,
but the solar metallicity well approximates the metallicity of
our samples. As long as the metallicity
is between 1/10 and 3 solar metallicity, the major source for
the uncertainty is the age, not the metallicity.

\subsection{Application to the {\it IUE} sample}
\label{subsec:result_IUE}

The {\it IUE} sample is also analysed in the way described in
Section \ref{subsec:result_SFG}. If we adopt the conversion
factors for $t_{\rm SF}=10^8$ yr, we obtain the results as
listed in Table \ref{tab:sb_sample}. For
some of the galaxies, $f$ is larger than 1, although $f<1$ by
definition. We suspect that this is due to the same reason as
CGCG 119054 (Section \ref{subsec:result_SFG}).
Indeed, the galaxies whose $f$ is larger than 1 have
a large $A({\rm H}\alpha )$ (1.42, 1.69, and 1.96 for IC 1586,
NGC 5860, and NGC 4194, respectively). We exclude those
galaxies from the following analysis.
The mean values ($\pm\sigma$) are $f=0.48$ ($\pm 0.20$),
$\epsilon =0.76$ ($\pm 0.15$), and $\eta =-0.04$ ($\pm 0.09$).
If we apply the conversion factors at $10^7$ yr (i.e., we adopt
$t_{\rm SF}=10^7$ yr), we obtain $f=0.30$ ($\pm 0.13$),
$\epsilon =0.76$ ($\pm 0.15$), and $\eta =-0.07$ ($\pm 0.09$).
As seen in the SFG sample, $f$ is sensitive to $t_{\rm SF}$
also for the {\it IUE}\, sample.

The extinctions for Lyman continuum and UV are both larger
for the {\it IUE}\, sample than for the SFG sample. For the
{\it IUE}\, sample, about 50\% of the Lyman continuum photons
and roughly 80\% of the UV photons are absorbed by dust grains.
Moreover almost all the dust heating source is the stellar
population younger than $10^8$ yr for the {\it IUE}\, sample,
since $\eta\sim 0$. This is consistent with the starburst
property of the {\it IUE}\, sample,
whose current star formation activity dominates the luminosity
of galaxies.

We examine the correlation between the quantities. There is no
evidence for the correlations between $f$ and $\epsilon$
($r=-0.12$) and between $f$ and $\eta$ ($=-0.04$). There seems to
be a tight relation
between $\epsilon$ and $\eta$, but this results from the same
reason as the SFG
sample (Section \ref{subsec:result_SFG}). In reality, the scatter
should be much larger if we consider the scatter in the
relation between IR/UV flux ratio and $\epsilon$.

\subsection{Validity of IHK}\label{subsec:validIHK}

One of our main aims is to examine the conversion formula of IHK
(eq.\ \ref{eq:conversion_formula}). In order to
see if the IHK formula works well or not, we should know the
best estimate for the SFR first of all. We have shown
that the SFR is traced very well by using IR and UV SFRs as
equation (\ref{eq:SFR_IR_UV}).
Therefore, in this paper the real SFR estimated observationally
is defined as
\begin{eqnarray}
{\rm SFR(best)}\equiv (1-\eta )\, C_{\rm IR}^{\rm sb}\,L_{\rm IR}
+C_{2000}\, L_{2000}\,\, .
\label{eq:realSFR}
\end{eqnarray}

\begin{figure*}
\begin{center}
\includegraphics[width=8cm]{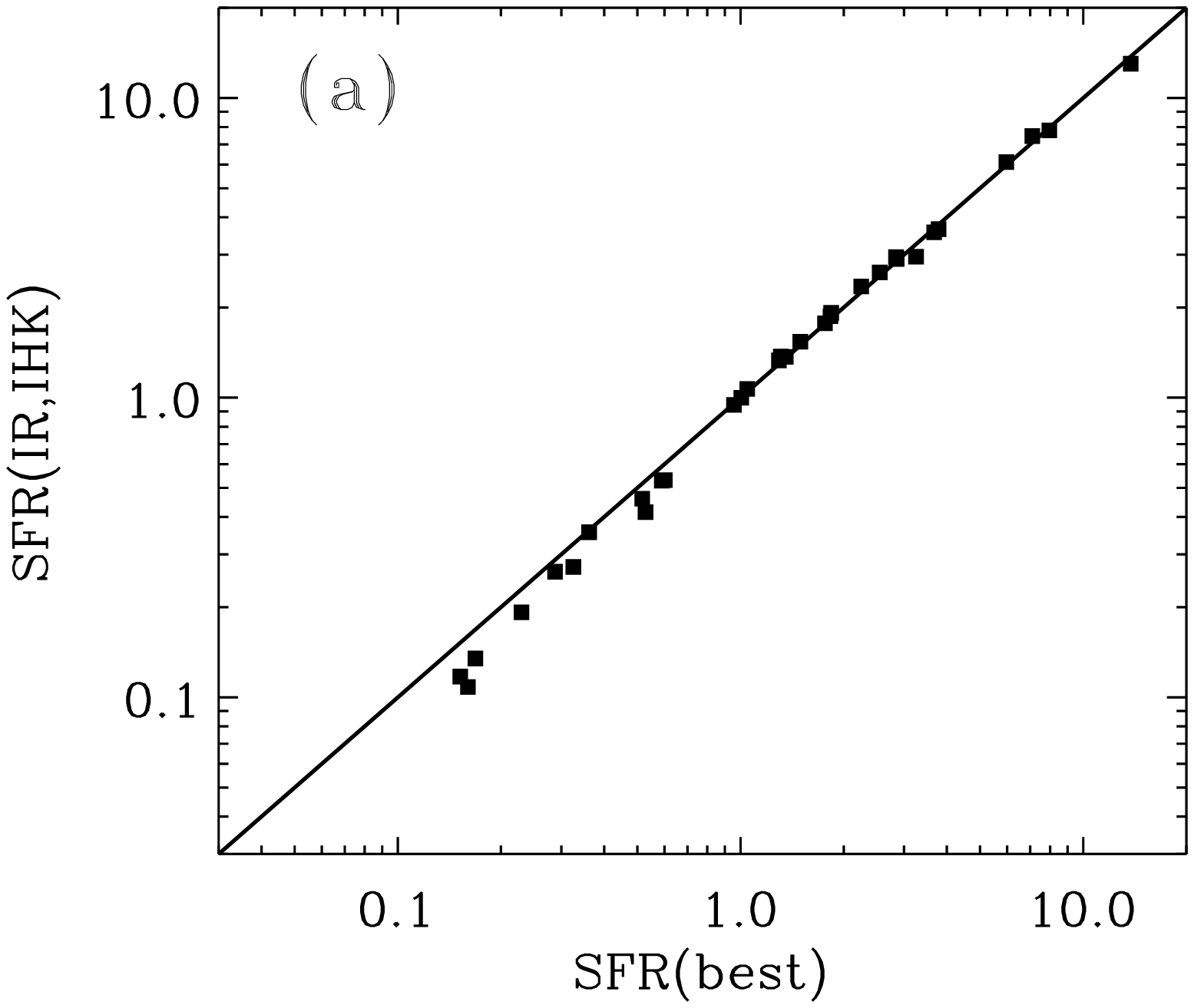}
\includegraphics[width=8cm]{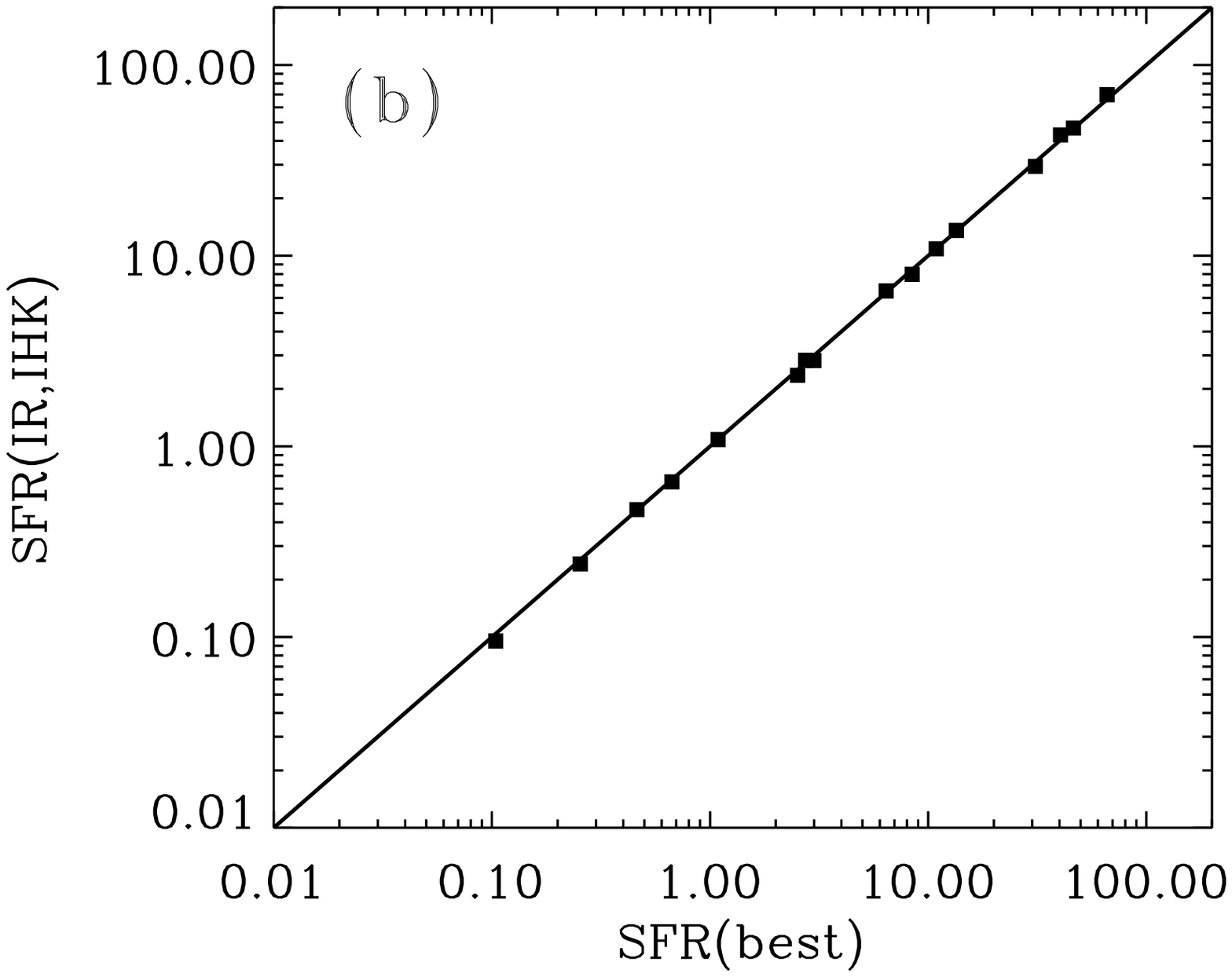}
\end{center}
\caption{Star formation rate derived from the IHK method,
SFR(IR,\,IHK), compared with the best-estimate SFR for the SFG
sample and the {\it IUE}\, sample (a and b, respectively).}
\label{fig:validIHK}
\end{figure*}

Since $f$, $\eta$ and $\epsilon$ are supposed to be known at this
step, we can examine the conversion factor for the IR SFR by
using the IHK method. We define the
following SFR(IR,\,IHK) by using the IHK conversion factor
(eq.\ \ref{eq:def_cir}):
\begin{eqnarray}
{\rm SFR(IR,\, IHK)}\equiv C_{\rm IR}(f,\,\epsilon ,\,\eta )\, 
L_{\rm IR}\, .
\end{eqnarray}
Using the quantities
$(f,\,\epsilon,\,\eta )$ derived for each galaxy in
Table~\ref{tab:sfg_sample}, we show the relation between
SFR(IR,\,IHK) and SFR(best) in Figure \ref{fig:validIHK}.
We find that SFR(IR,\,IHK) agrees with SFR(best) within
a difference of $\sim 30$\% (for 70\% of
the SFG sample, the difference is within 10\%). Therefore,
if we know the three quantities, the IHK method approximates
the SFR very well.

For a general sample, we do not necessarily have all the three
(UV, H$\alpha$, and dust IR) luminosities. In this case,
$C_{\rm IR}(f,\,\epsilon ,\,\eta )$ cannot be obtained by
our method. Thus, in the next subsection, we examine
various SFRs derived from a limited number of luminosities.

\subsection{SFR}\label{subsec:sfr_comp}

\begin{figure*}
\begin{center}
\includegraphics[width=8cm]{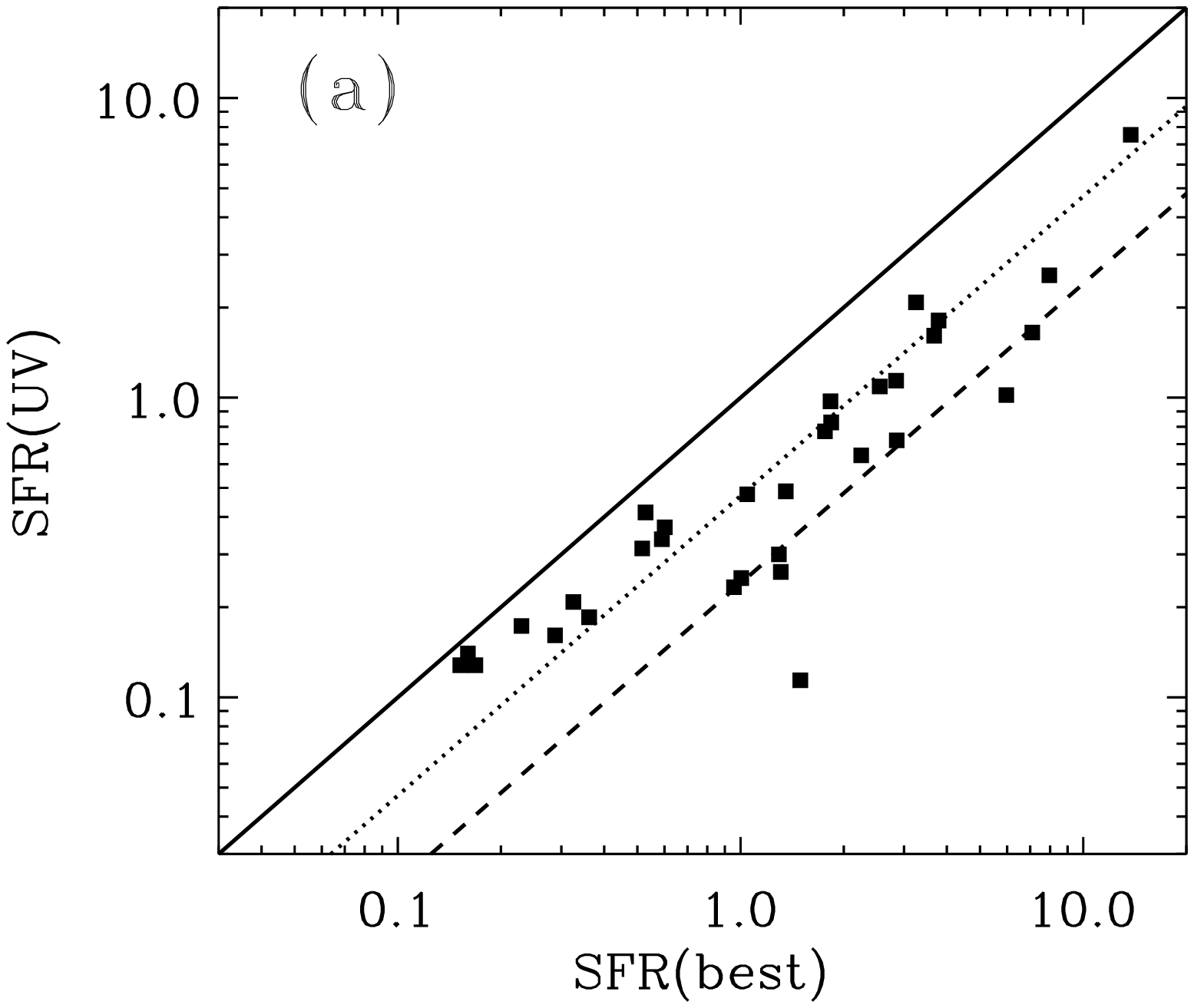}
\includegraphics[width=8cm]{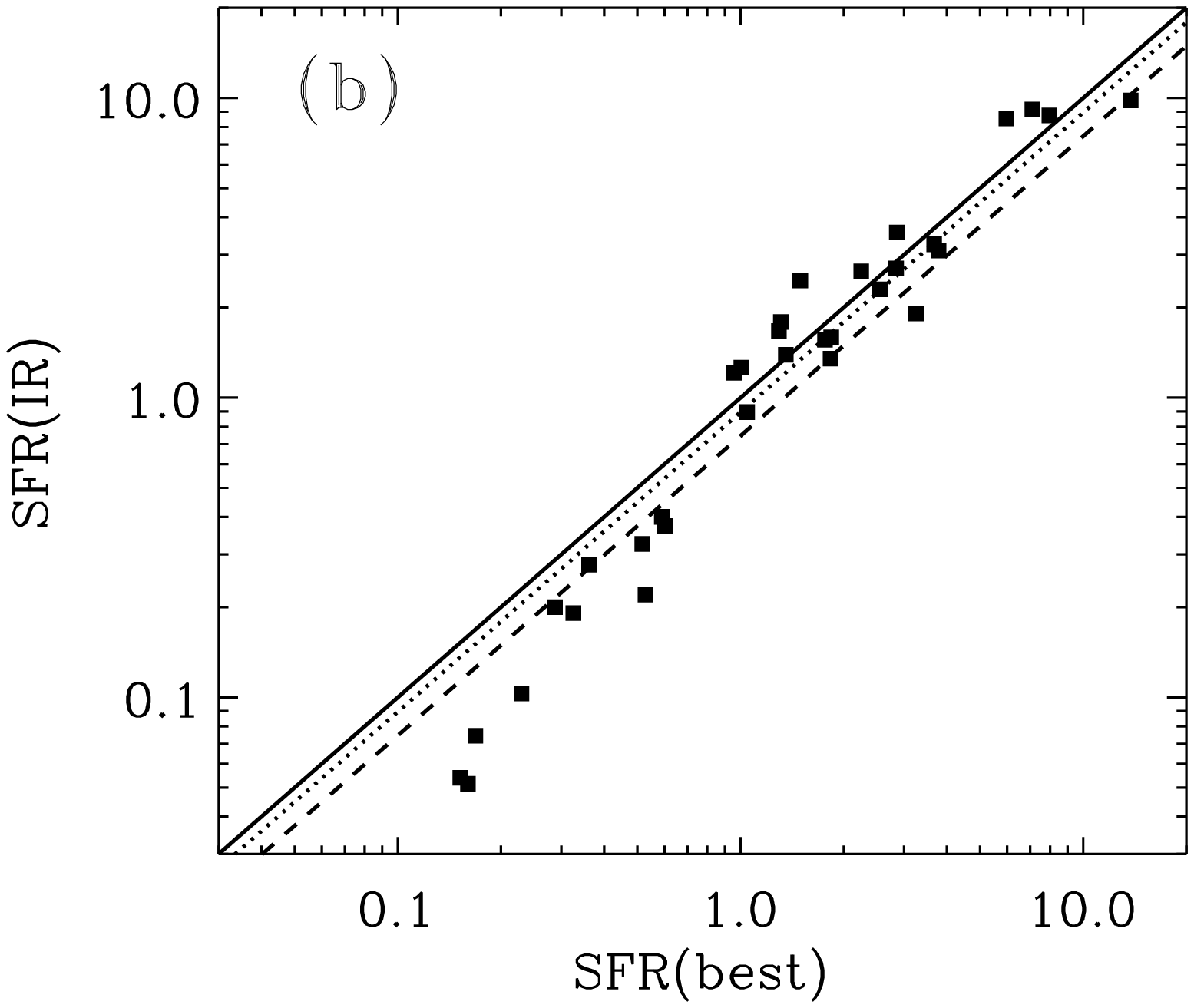}\\
\includegraphics[width=8cm]{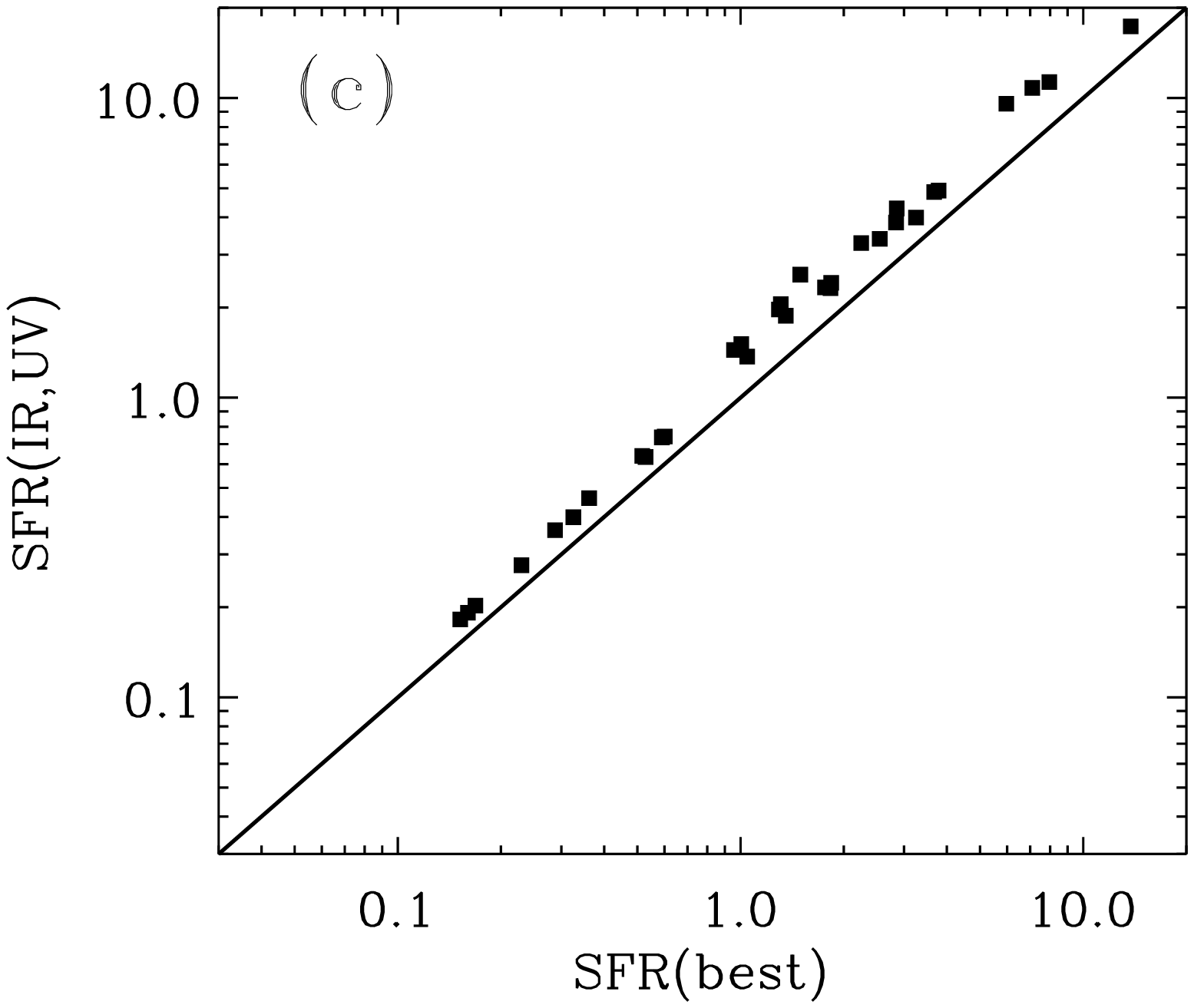}
\includegraphics[width=8cm]{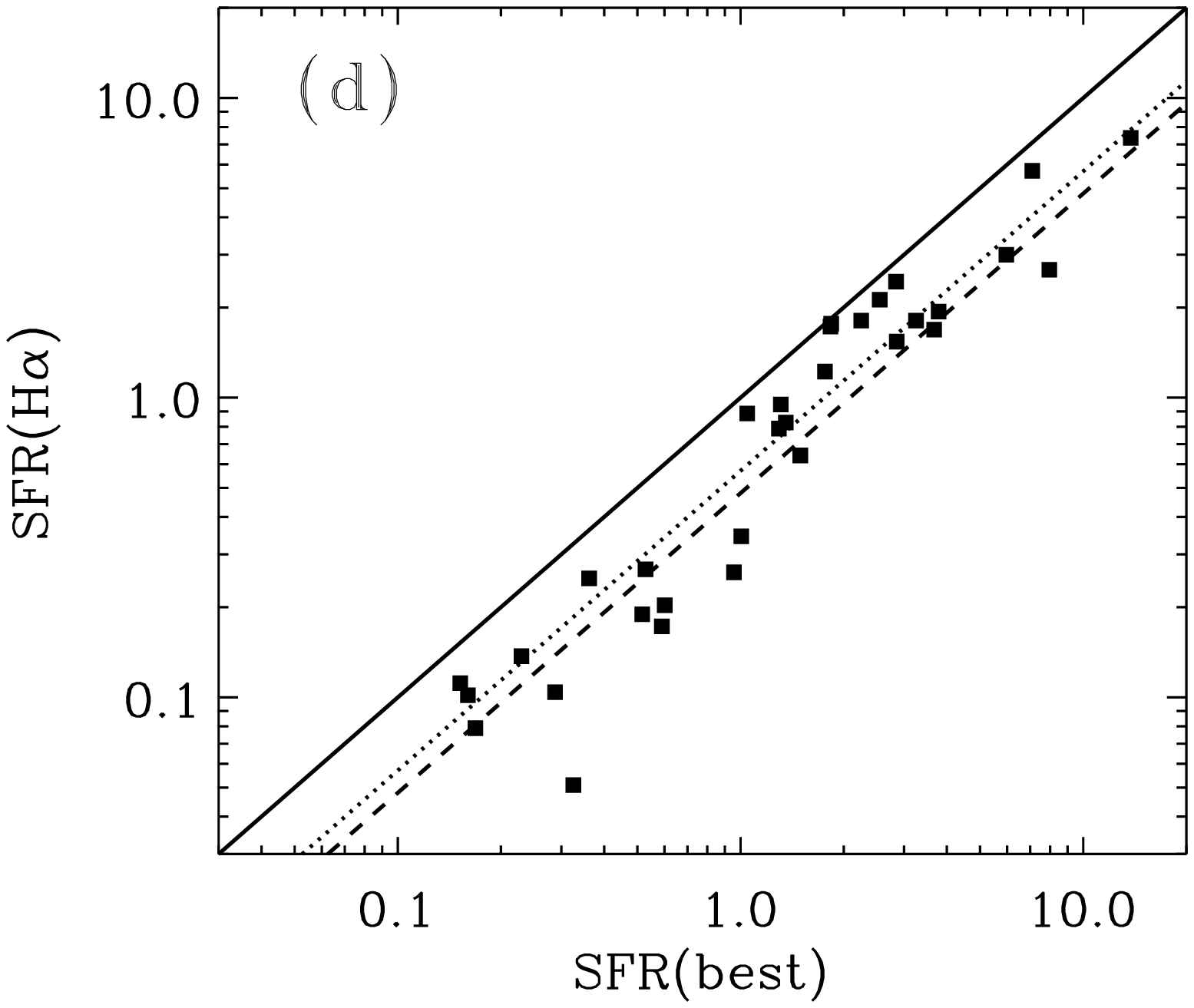}
\end{center}
\caption{Various SFRs of the SFG sample plotted against the
best-estimate SFR. The solid lines show the equality of the two
SFRs plotted in each figure. (a) SFR(UV) (without any
extinction correction).
The dotted and dashed lines indicate the ``extinction-corrected''
relations, SFR(UV)/$(1-\epsilon )={\rm SFR(best)}$ with
$\epsilon =0.53$ (the mean of the SFG sample) and with
$\epsilon =0.76$ (the mean of the {\it IUE}\, sample), respectively.
(b) SFR(IR) converted directly from the dust IR luminosity by
multiplying $C_{\rm IR}^{\rm sb}$. The solid line shows
${\rm SFR(IR)}={\rm SFR(best)}$. The dotted and dashed lines
present the trend for the SFG sample and the {\it IUE}\, sample,
respectively. (c)
${\rm SFR(IR,\,UV)}\equiv{\rm SFR(UV)}+{\rm SFR(IR)}$.
(d) SFR(H$\alpha$) derived from the H$\alpha$ luminosity corrected
for the Balmer decrement. The dotted and dashed lines show the
relations corrected for the Lyman continuum extinction with
$f=0.57$ (the mean of the SFG sample) and $f=0.48$ (the mean of the
{\it IUE}\, sample, respectively.}
\label{fig:sfr_compare}
\end{figure*}

Here we investigate SFRs derived from various indicators. In
Figures~\ref{fig:sfr_compare}a--d, we compare SFR(UV), SFR(IR),
SFR(IR,\,UV), and SFR(H$\alpha$) with SFR(best) for
the SFG sample. Figure \ref{fig:sfr_compare}a clearly shows that
SFR(UV) underestimates the SFR because of dust absorption. The
ratio SFR(UV)/SFR(best) is equal to
$(1-\epsilon )$ (eq.\ \ref{eq:SFR_corr_UV}). If we can estimate
$\epsilon$ (UV extinction) for each galaxy, SFR(UV)/$(1-\epsilon )$
gives the best estimate of SFR.

The discrepancy between SFR(UV) and SFR(best)
tends to be small for small SFR. For
${\rm SFR(best)}\la 1~\solyr$, SFR(UV) gives a good estimate for
the real SFR (see also
Bell \& Kennicutt \cite{bell01}). This means that there is a
positive correlation between SFR and $\epsilon$
as pointed out also by Hopkins et al.\ (\cite{hopkins01}).
This correlation may only reflect the size effect, since a large
galaxy may tend to contain a lot of star-forming regions and at
the same time a large optical depth of dust
(Wang \& Heckman \cite{wang96}; Buat \& Burgarella \cite{buat98}).

We expect that ${\rm SFR(UV)}/(1-\epsilon )$ with $\epsilon =0.53$
gives a better estimate for the SFR. The dotted line in
Figure \ref{fig:sfr_compare}a shows the relation
${\rm SFR(UV)}/(1-\epsilon )={\rm SFR(best)}$ with $\epsilon=0.53$.
We observe that ${\rm SFR(UV)}/(1-\epsilon )$ with $\epsilon =0.53$
systematically overestimates the SFR for
${\rm SFR}\la 1~\solyr$, because the data points lie in the region
${\rm SFR(UV)}/(1-\epsilon )>{\rm SFR(best)}$. Thus, the dust
correction should be varied depending on the SFR. Because of a
large variety in $\epsilon$, the scatter of
SFR(UV) is larger than that of any other estimators.
The dashed line in Figure \ref{fig:sfr_compare} represents
the relation applied to the {\it IUE}\, sample (i.e.,
$\epsilon =0.76$).

Figure \ref{fig:sfr_compare}b indicates that SFR(IR) estimates
the SFR quite well for
${\rm SFR}\ga 1~\solyr$. However, we should keep in mind that this
is the result of the cancellation of the following under- and
overestimate (Kennicutt \cite{kennicutt98a};
Inoue \cite{inoue02a}): SFR(IR) overestimates the SFR by a factor
$1/(1-\eta )$ because a part of the IR dust luminosity
originates from the old stellar population; SFR(IR) underestimates
the SFR because a part of the radiation originating from young
stars is not absorbed by dust, and thus is not traced by
IR. However, SFR(IR)
systematically underestimates SFR(best) when the SFR is
lower than 1 $\solyr$, because the major part of the energy is
radiated in UV. Accordingly, SFR(UV) provides a reasonable
estimate of SFR for ${\rm SFR}\la 1~\solyr$.

The conversion factor for the IR luminosity can be tested by
using the IHK conversion factor
$C_{\rm IR}(f,\,\epsilon,\,\eta )$. In treating a sample of
galaxies, a typical values
of $f$, $\epsilon$, and $\eta$, for example
the mean values ($f=0.57$, $\epsilon =0.53$, and
$\eta =0.40$ for the SFG sample), are useful. If we put those
mean values, we find
that $C_{\rm IR}=2.0\times 10^{-10}~\solyr~L_\odot^{-1}$. This
value is similar to
$C_{\rm IR}^{\rm sb}=1.79\times 10^{-10}~\solyr~L_\odot^{-1}$.
If we use $2.0\times 10^{-10}~\solyr~L_\odot^{-1}$ instead of
$C_{\rm IR}^{\rm sb}$ to estimate SFR(IR), the data points in
Figure \ref{fig:sfr_compare}b shift upwards. In order not to
complicate the figure, we shift the solid line down to the
dotted line. The dashed line shows the same thing for
$C_{\rm IR}=2.4\times 10^{-10}~\solyr~L_\odot^{-1}$, which
is representative for the {\it IUE}\, sample
(Table \ref{tab:formula}).
If $f$, $\epsilon$, and $\eta$ move their 1 $\sigma$ ranges,
$C_{\rm IR}=1.3$ -- $3.5\times 10^{-10}~\solyr~L_\odot^{-1}$.
For a comparison, we should note that Buat \& Xu (\cite{buat96})
derive a similar range
$C_{\rm IR}=0.79$ -- $2.6\times 10^{-10}~\solyr~L_\odot^{-1}$,
where we assume $L_{\rm IR}/L_{\rm FIR}=2.4$ (the mean for the
SFG sample).

\begin{table*}
\begin{center}
\caption{Recommended conversion factors from each luminosity
to SFR on the timescale of $t_{\rm SF}=10^8$ yr.}
\begin{tabular}{p{0.15\linewidth}ccl}\hline\noalign{\smallskip}
Luminosity$^{\rm a}$ &
\multicolumn{2}{c}{Multiplying factor$^{\rm b}$} & Comment \\
 & normal$^{\rm c}$ & starburst$^{\rm d}$ & \\
 \noalign{\smallskip}\hline\noalign{\smallskip}
$L_{2000}$ & $4.3\times 10^{-40}$ & $8.5\times 10^{-40}$ &
 large dispersion in extinction \\
 & & & systematic underestimate for $\la 1~\solyr$ \\
 \noalign{\smallskip}
$L_{\rm H\alpha}^{\rm c}$ & $1.4\times 10^{-41}$ & $1.6\times 10^{-41}$
& similar factor for both (``universal'') \\
 & & & applicable to any SFR \\
 \noalign{\smallskip}
$L_{\rm IR}$ & $2.0\times 10^{-10}$ & $2.4\times 10^{-10}$ &
 risk of underestimate for $\la 1~\solyr$ \\
\noalign{\smallskip}
\hline\noalign{\smallskip}
\multicolumn{4}{c}{Other formulae}\\ \noalign{\smallskip}
\multicolumn{3}{c}{$(1-\eta )\, C_{\rm IR}^{\rm sb}\, L_{\rm IR}+
 C_{2000}\, L_{2000}$} & $\eta$ necessary \\
\multicolumn{3}{c}{$C_{\rm IR}^{\rm sb}\, L_{\rm IR}+C_{2000}\, L_{2000}$}
 & systematic overestimate for normal galaxies \\
\noalign{\smallskip}
\hline
\end{tabular}
\begin{list}{}{}
\item[$^{\mathrm{a}}$] $L_{\rm H\alpha}^{\rm c}$ is the H$\alpha$
luminosity after the correction for the H$\alpha$ extinction
($A({\rm H}\alpha )$). $L_{\rm IR}$ is the total luminosity of
dust emission derived from {\it IRAS}\, 40--120 $\mu$m luminosity
and 60 $\mu$m vs.\ 100 $\mu$m flux ratio
(Dale et al.\ \cite{dale01}).
\item[$^{\mathrm{b}}$] The units are the same as in
Table \ref{tab:notation}.
\item[$^{\mathrm{c}}$] The SFG sample is assumed to be representative
of normal star-forming galaxies.
\item[$^{\mathrm{d}}$] The {\it IUE}\, sample is assumed to be
representative of starburst galaxies.
\end{list}
\label{tab:formula}
\end{center}
\end{table*}

We also examine the following SFR defined as a simple sum of
IR and UV SFRs:
\begin{eqnarray}
{\rm SFR(IR,\, UV)}\equiv C_{\rm IR}^{\rm sb}\, L_{\rm IR}+C_{2000}
\, L_{2000}\, .\label{eq:iruvSFR}
\end{eqnarray}
This kind of sum is adopted in
Flores et al.\ (\cite{flores99}) and Buat et al.\ (\cite{buat99}).
In Figure \ref{fig:sfr_compare}c, we show the relation between
SFR(IR,\,UV) and SFR(best). We observe that SFR(IR,\,UV)
overestimates the SFR because the fraction $\eta$ related to
the old stars is not subtracted from $L_{\rm IR}$. However, the
overestimate is not so large, $\sim 60\%$ at most. Moreover,
the systematic SFR-dependent deviation, which is seen for
SFR(UV) and SFR(IR), disappears by the combination of UV and IR
SFRs.
If we know a typical $\eta$ for a sample of galaxies, it is
possible to statistically subtract the contribution from old
stars by using
equation (\ref{eq:realSFR}). 

The H$\alpha$ SFR defined by the following expression is also
tested:
\begin{eqnarray}
{\rm SFR(H\alpha)}\equiv C_{\rm H\alpha}\, L_{\rm H\alpha}^{\rm c}
\, .
\end{eqnarray}
SFR(H$\alpha$) is plotted against SFR(best) in
Figure \ref{fig:sfr_compare}d. Because $f$ is significantly smaller
than 1, SFR(H$\alpha$) underestimates the SFR. The dotted line
in Figure \ref{fig:sfr_compare}d shows the relation
${\rm SFR(H\alpha )}/f={\rm SFR(best)}$ with $f=0.57$ (the mean 
value). This
line reproduces the mean trend of the data over all the
range of SFR. This trend strongly supports the usefulness of
H$\alpha$ luminosity as an indicator of SFR, because the
H$\alpha$ luminosity is independent of SFR(best), while
SFR(best) includes dependence on SFR(UV) and SFR(IR)
(thus the two SFRs in each of the figures a--c are not
fully independent). Therefore, we conclude that
${\rm SFR(H\alpha )}/f$
with $f=0.57$ gives a good estimate for SFR of the star-forming
galaxies over the wide range of SFR.
The dispersion in the figure is produced partly due to the different 
age in
the present star formation activity, because H$\alpha$ traces
the star formation in recent $10^{6}$--$10^7$ yr while UV and
FIR traces all the star formation activity in recent $\sim 10^8$
yr or more.

We should keep in mind that we have adopted the H$\alpha$ luminosity
corrected for $A({\rm H}\alpha )$. The analysis by
Hopkins et al.\ (\cite{hopkins01}) suggests that $A({\rm H}\alpha )$
correlates with SFR (see also B02). Therefore, the conversion factor
for observed H$\alpha$
luminosity before the correction for $A({\rm H}\alpha )$ is not
universal but dependent on the SFR.
It is also important that we should take into account not only
$A({\rm H}\alpha )$ but also the Lyman continuum extinction in
order to obtain a reliable
estimate of the SFR. If we do not correct for the Lyman continuum
extinction, the SFR is systematically
underestimated by a factor of $\sim 2$
(see also Inoue et al.\ \cite{inoue01}).

\begin{figure*}
\begin{center}
\includegraphics[width=8cm]{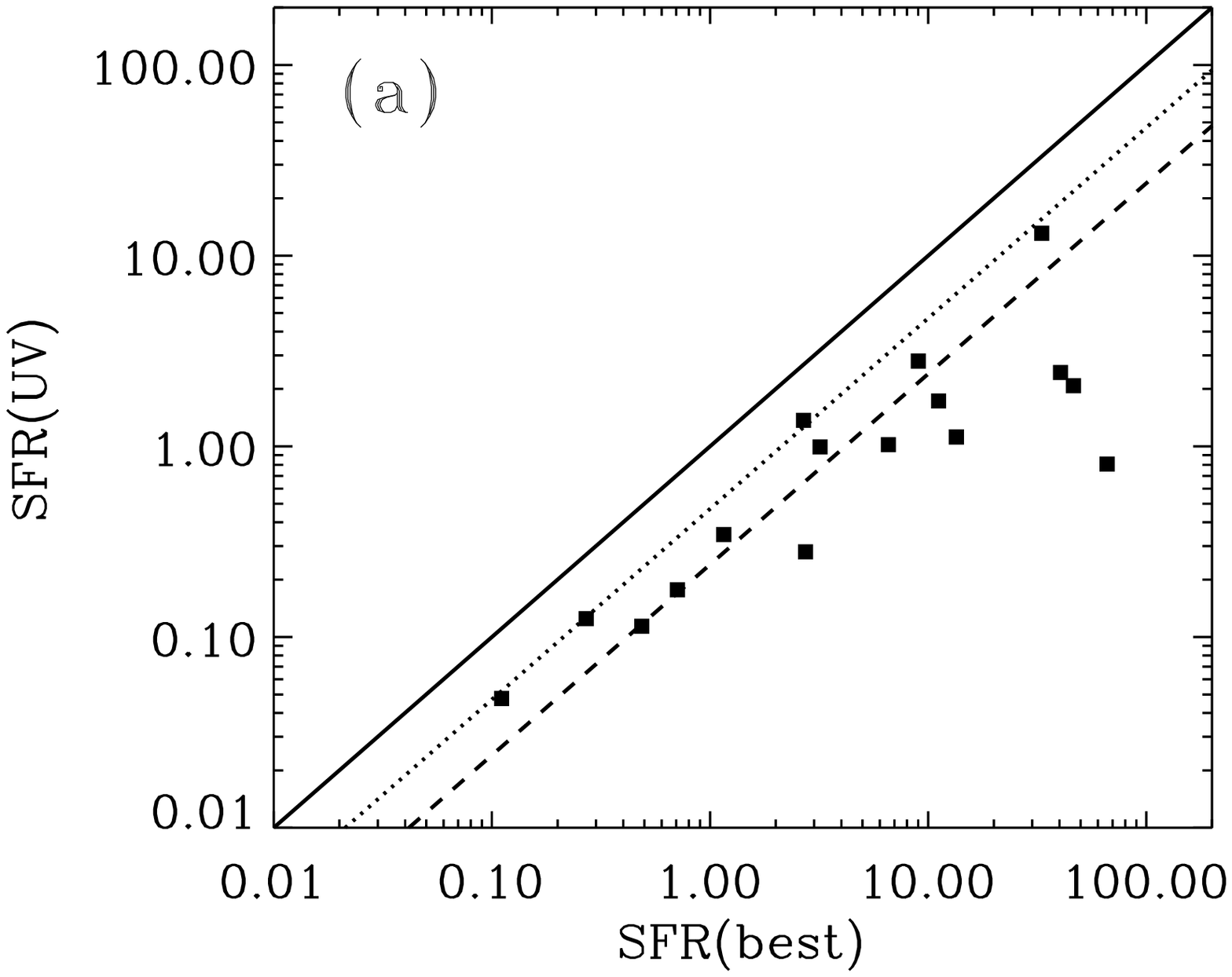}
\includegraphics[width=8cm]{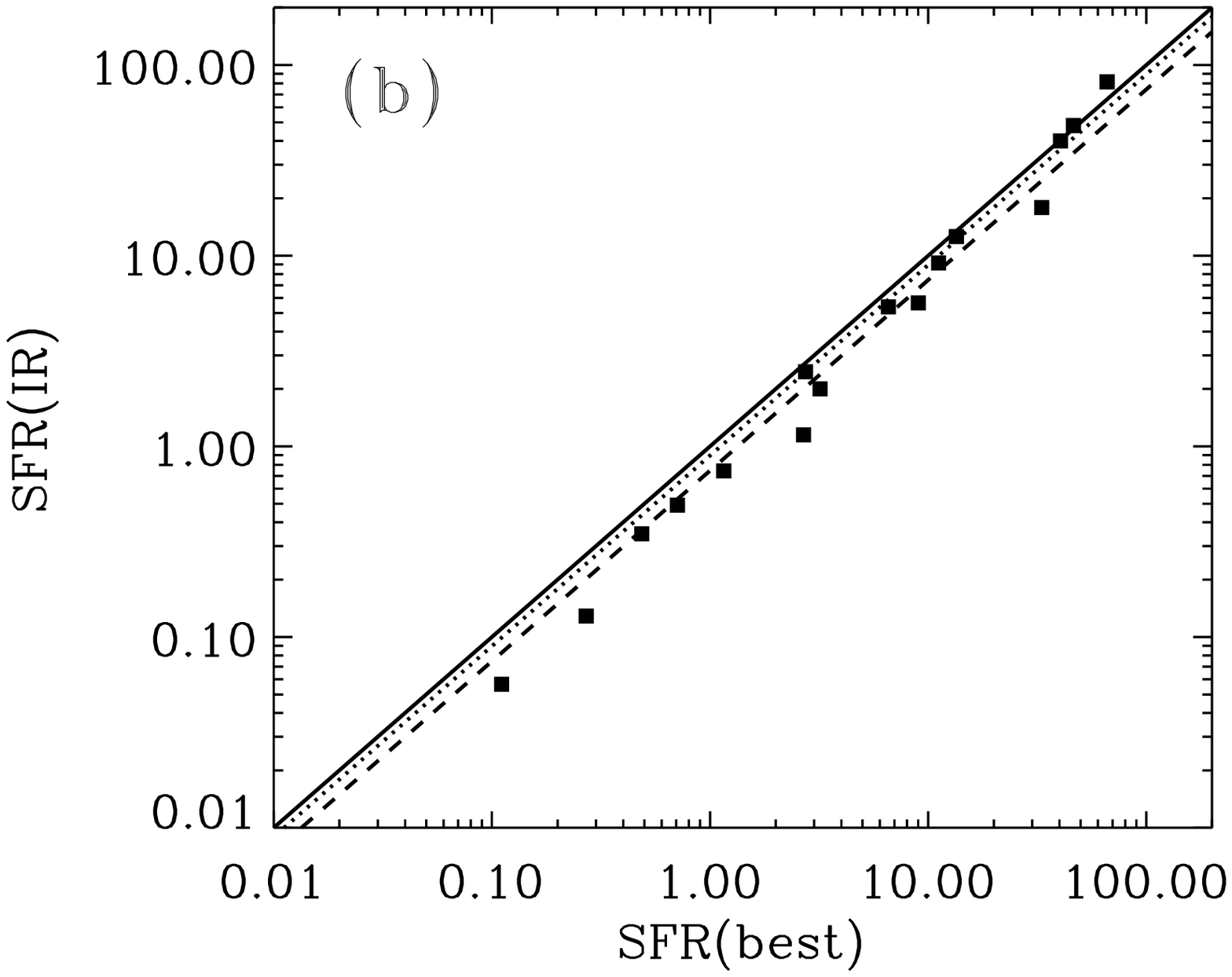}\\
\includegraphics[width=8cm]{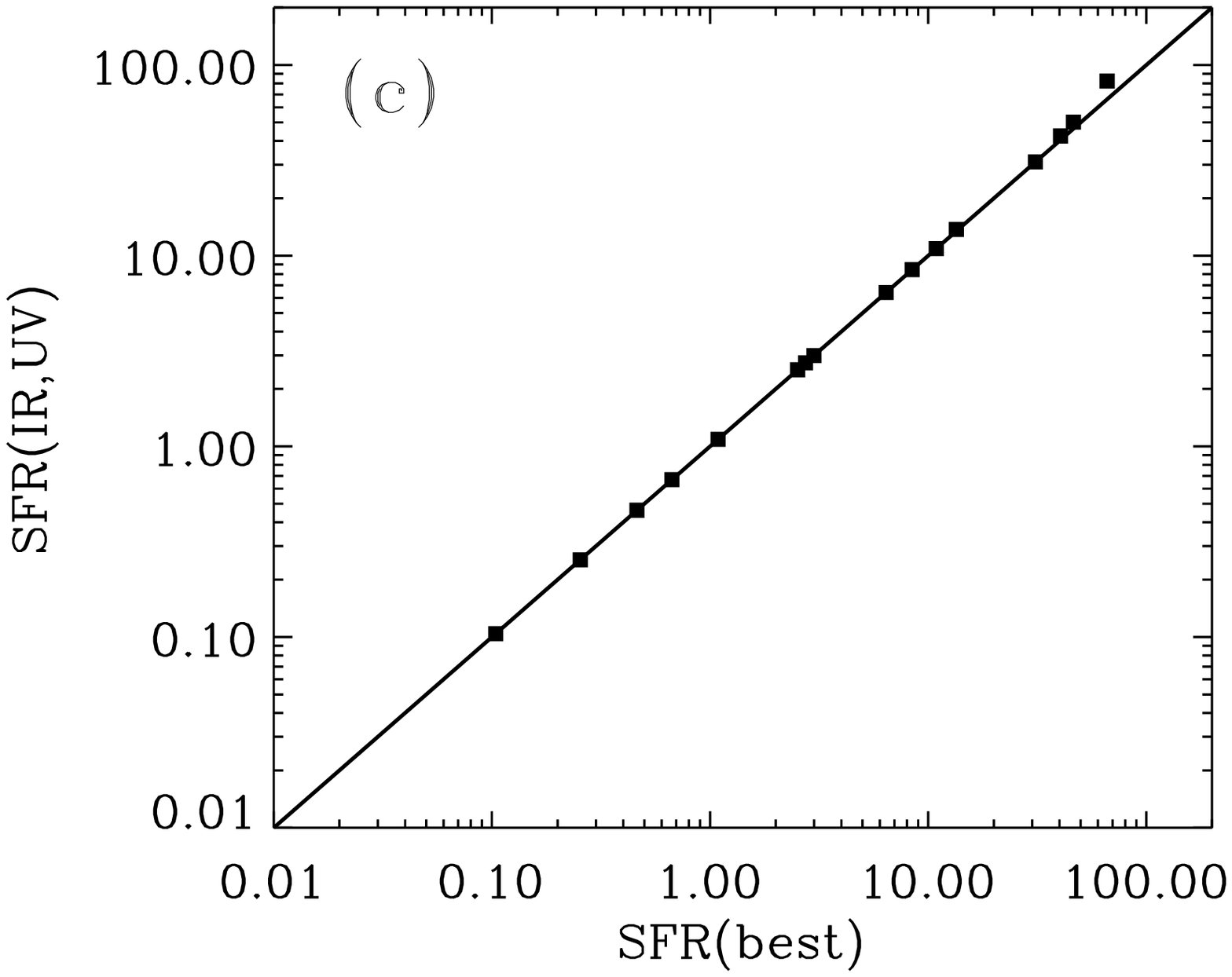}
\includegraphics[width=8cm]{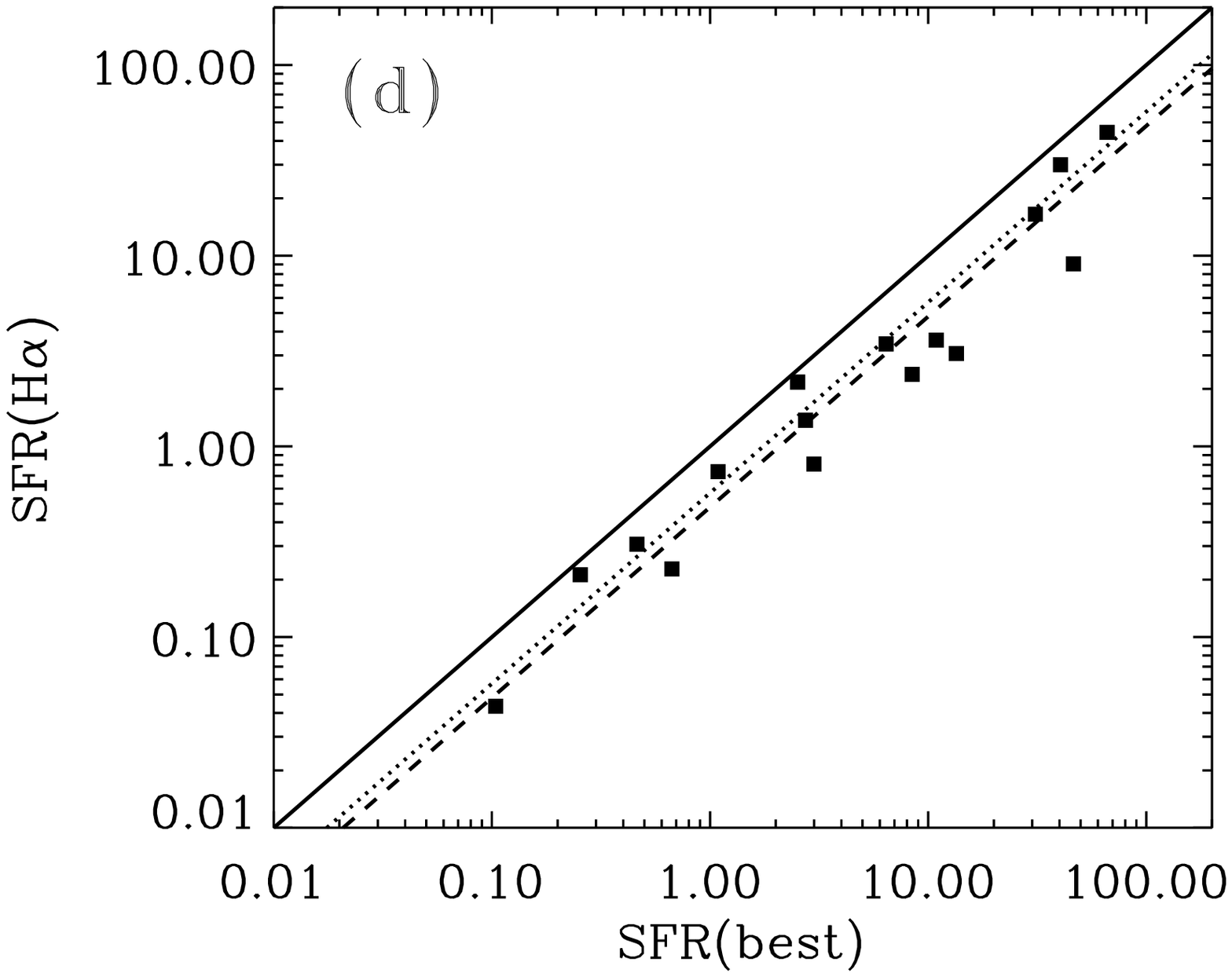}
\end{center}
\caption{Same as Fig.\ \ref{fig:sfr_compare} but for the
{\it IUE}\, sample. If $\eta <0$, we adopt $\eta =0$.}
\label{fig:sfrsb_compare}
\end{figure*}

In Figure \ref{fig:sfrsb_compare}, we examine the {\it IUE}\,
sample. Since $\eta <0$ is not allowed by definition, we assume
$\eta =0$ if $\eta <0$. This does not affect the following
discussions because $|\eta|\ll 1$ for the {\it IUE}\, sample.
We use the conversion factor for $t_{\rm SF}=10^8$ yr
also for this sample. In Figure \ref{fig:sfrsb_compare}a,
we show the line ${\rm SFR(UV)}/(1-\epsilon )={\rm SFR(best)}$
with $\epsilon =0.53$ (the mean value for the SFG sample) by the
dotted line. Since the {\it IUE}\, sample is much obscured in
UV, the correction with $\epsilon =0.53$ underestimates the
SFR. We also
show ${\rm SFR(UV)}/(1-\epsilon )={\rm SFR(best)}$
with $\epsilon =0.76$ (the mean value for the {\it IUE} sample)
by the dashed line, which fits the data points better.
Therefore, when we correct the UV SFR of a galaxy for dust 
absorption, it is necessary
to know if the galaxy is to be classified as a
normal star-forming galaxy or a starburst.
In this sense, there is no universal
correction factor for the UV SFR.

Figures \ref{fig:sfrsb_compare}b shows that SFR(IR)
approximates the SFR very well. This is because a large
fraction of stellar light is absorbed by dust and reprocessed in
IR in starburst galaxies. 
If we apply the mean values ($f=0.48$, $\epsilon =0.76$, and
$\eta =0$), we obtain
$C_{\rm IR}=2.4\times 10^{-10}~\solyr~L_\odot^{-1}$, a value
similar to $C_{\rm IR}^{\rm sb}$
(i.e., ${\rm SFR(IR,\,IHK)/SFR(IR)}=1.3$).
This is the reason
why SFR(IR) gives a good estimate for the SFR of the
starburst galaxies.
The range expected from the 1 $\sigma$ variations of
$f$, $\epsilon$, and $\eta$ is
$C_{\rm IR}=2.0$--$3.0\times 10^{-10}~\solyr~L_\odot^{-1}$.

In Figure \ref{fig:sfrsb_compare}c, we plot SFR(IR,\,UV) against
SFR(best) for the {\it IUE}\, sample. Since $\eta\simeq 0$
for the {\it IUE}\,
sample, SFR(IR,\,UV) is almost equal to the best-estimate SFR
(compare equations \ref{eq:realSFR} and \ref{eq:iruvSFR}).
Therefore, for starburst galaxies, the simple sum of the UV and
IR SFRs gives the best estimate of SFR.

The H$\alpha$ SFR of the {\it IUE}\, sample is also shown in
Figure \ref{fig:sfrsb_compare}d, in which the dashed line shows
the relation ${\rm SFR(H\alpha )}/f={\rm SFR(best)}$ with $f=0.48$
(the mean value for the {\it IUE}\, sample). The dotted line
presents the same relation with $f=0.57$ (the mean for the SFG
sample). The dashed line reproduces the mean trend of the data
over all the
range of SFR. Moreover, there is little difference between the
dotted and dashed lines in Figure \ref{fig:sfr_compare}e, which
means that the same conversion factor can be used for H$\alpha$
luminosity. Therefore, we suggest that ${\rm SFR(H\alpha )}/f$
with $f\sim 0.5$ gives a good estimate for SFR of both the SFG
and {\it IUE} samples. 
Unlike the UV and IR conversion factors, there is no systematic
difference in the H$\alpha$ conversion factor in all the SFR
range.

We summarise the above various ways of SFR estimate in
Table \ref{tab:formula}, where we list the conversion
factors applicable to both types of galaxies. We assume that
the SFG sample is representative of normal star-forming
galaxies and that the {\it IUE}\, sample is representative of
starburst galaxies.
The listed conversion factors are already corrected for dust
effects (except for the H$\alpha$ absorption,
$A({\rm H}\alpha)$, which should
be estimated independently by Balmer decrement) and are our
``recommended'' values.

\subsection{Metallicity}\label{subsec:metallicity}

H01 suggest that $f$ and $\epsilon$ change as a function of
metallicity because the optical depth of dust can be related
to dust-to-gas ratio (eqs.\ 11 and 13 of H01) and
dust-to-gas ratio increases as metallicity increases. We do not
show the relation between $\eta$ and metallicity in this paper
because it only shows $\eta\sim 0.4$ for
the SFG sample and $\eta\simeq 0.0$ for the {\it IUE}\, sample
with a small scatter and without any correlation with
metallicity. The scatter of $\eta$ should be larger in reality
if we consider the scatter between the relation between
IR/UV flux ratio and $\epsilon$
(Section \ref{subsec:result_SFG}).

\begin{figure*}
\begin{center}
\includegraphics[width=8cm]{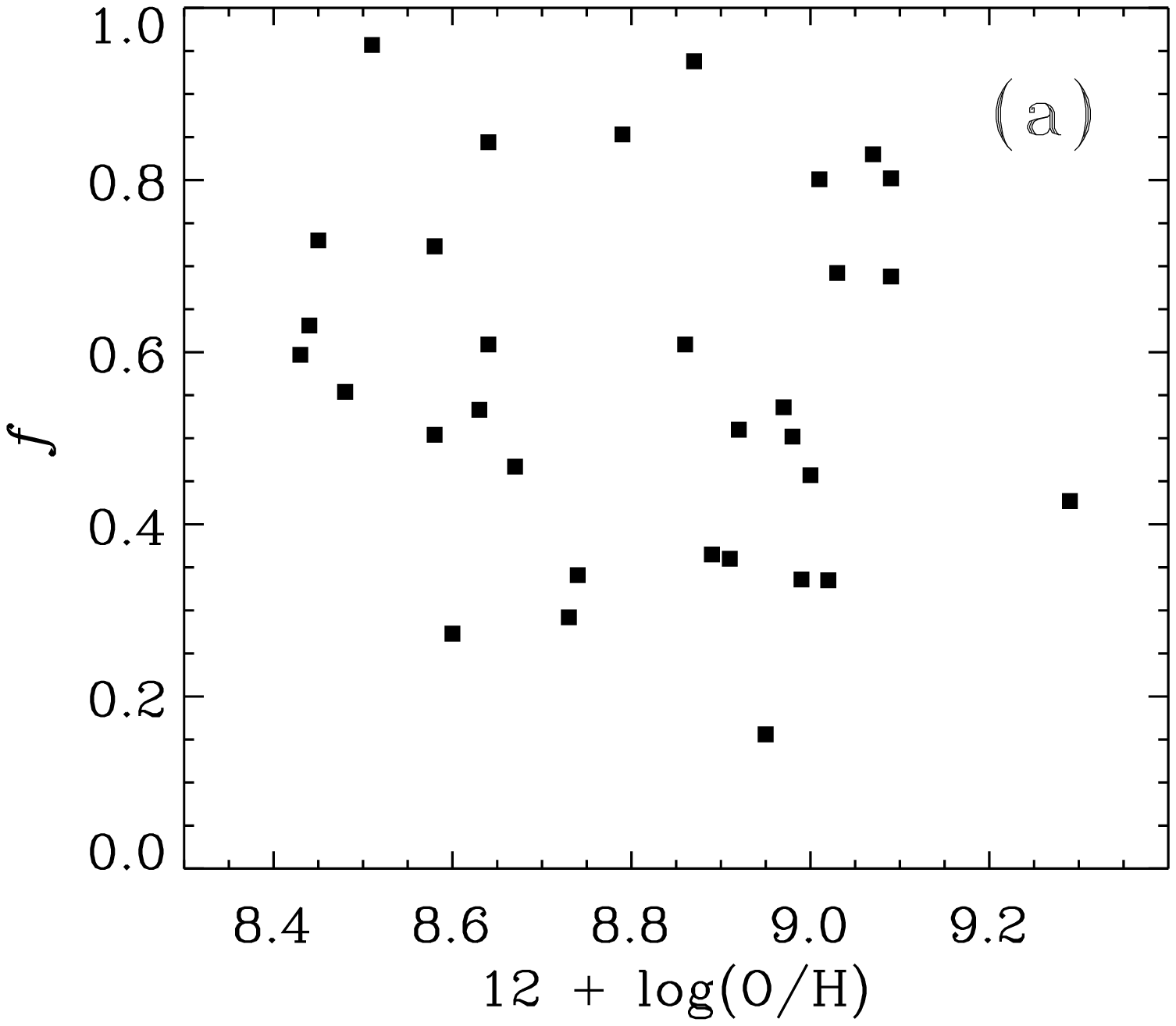}
\includegraphics[width=8cm]{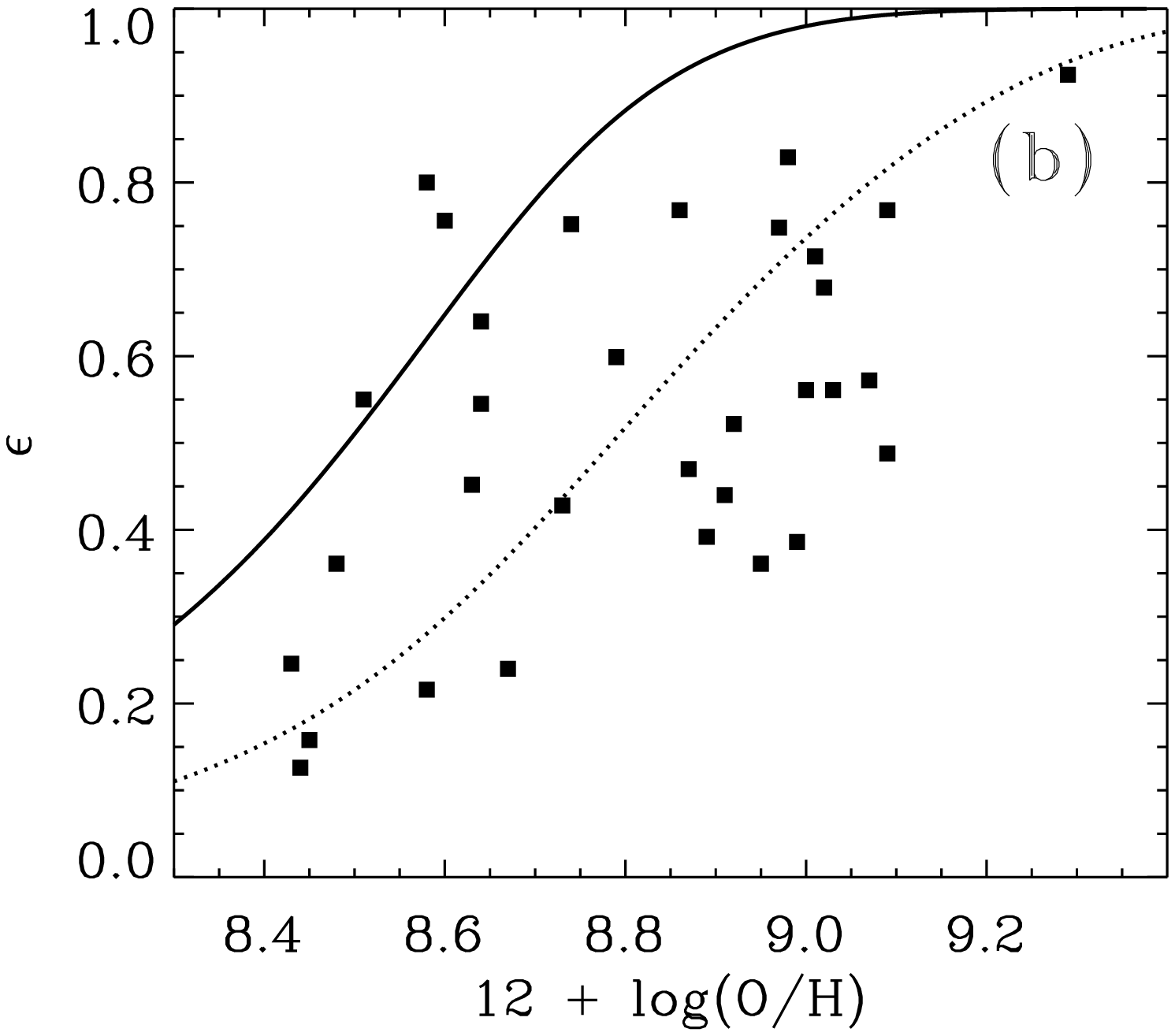}
\end{center}
\caption{(a) $f$ vs.\ metallicity, and (b) $\epsilon$ vs.\
metallicity for the SFG sample. The solid
line in (b) shows the model by H01. The dotted line
represents the
result with our new normalisation for the optical depth of
the SFG sample.}
\label{fig:metal_sfg}
\end{figure*}

In Figure \ref{fig:metal_sfg}, we show the relation between
(a) $f$ and metallicity, and (b) $\epsilon$ and metallicity for
the SFG sample. In Figure \ref{fig:metal_sfg}a, we do not find
evidence for correlation ($r=-0.13$). There is a significant scatter
over the range of $0<f<1$. However,
Inoue et al.\ (\cite{inouehk01}) and
Inoue (\cite{inoue01}) show that
there is a correlation between $f$ and dust-to-gas ratio
for
H\,{\sc ii} regions. They as well as H01 estimate the optical
depth of dust for ionising photons by using
the Str\"{o}mgren sphere modelling of an H\,{\sc ii} region. The
resulting optical depth becomes a function of the gas density of
the H\,{\sc ii} region and the ionising photon luminosity
of the central star as well as dust-to-gas ratio
(Spitzer \cite{spitzer78}). Therefore, a possible interpretation
on the large scatter of $f$ is that the gas density and/or the
number of ionising photons per H\,{\sc ii} region differ
from galaxy to galaxy. H01 also assume that there is a tight
relation between dust-to-gas ratio and metallicity.
It is also shown that the scatter of dust-to-gas ratio can
be so large that the correlation between dust-to-gas ratio and
metallicity becomes weak (e.g., Hirashita et al.\
\cite{hirashita02}). This would make the correlation between $f$
and metallicity weak, even if there is a correlation between $f$
and dust-to-gas ratio. Another reason for the large scatter is
the presence of ionising photons escaping from H\,{\sc ii}
regions, which cannot be treated by the H\,{\sc ii} region
modelling. The variety of geometry of dust distribution 
(e.g., a dust-depleted region in the central parts of
H\,{\sc ii} regions; Inoue \cite{inoue02b}) can cause the large
scatter of $f$.

\begin{figure*}
\begin{center}
\includegraphics[width=8cm]{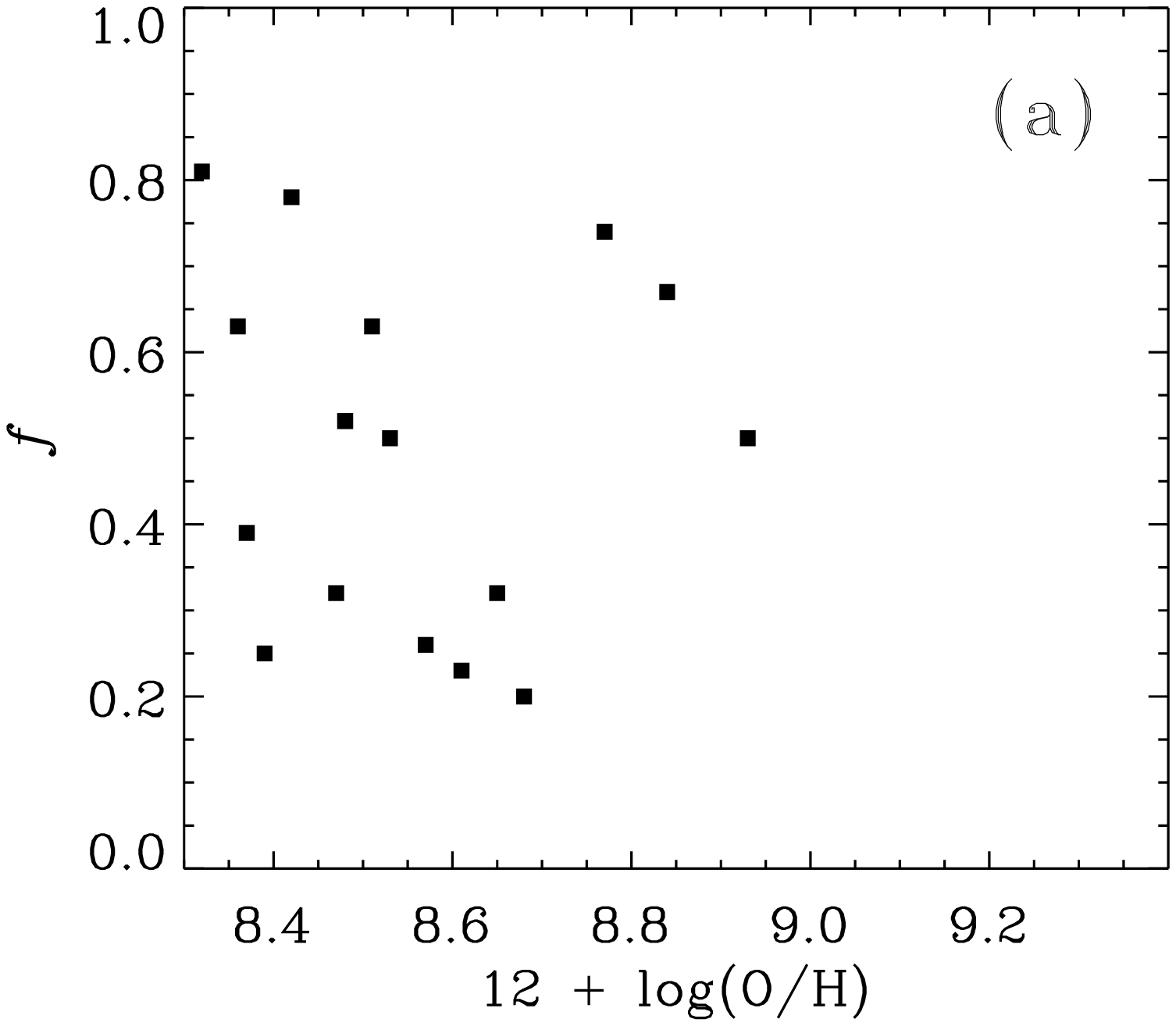}
\includegraphics[width=8cm]{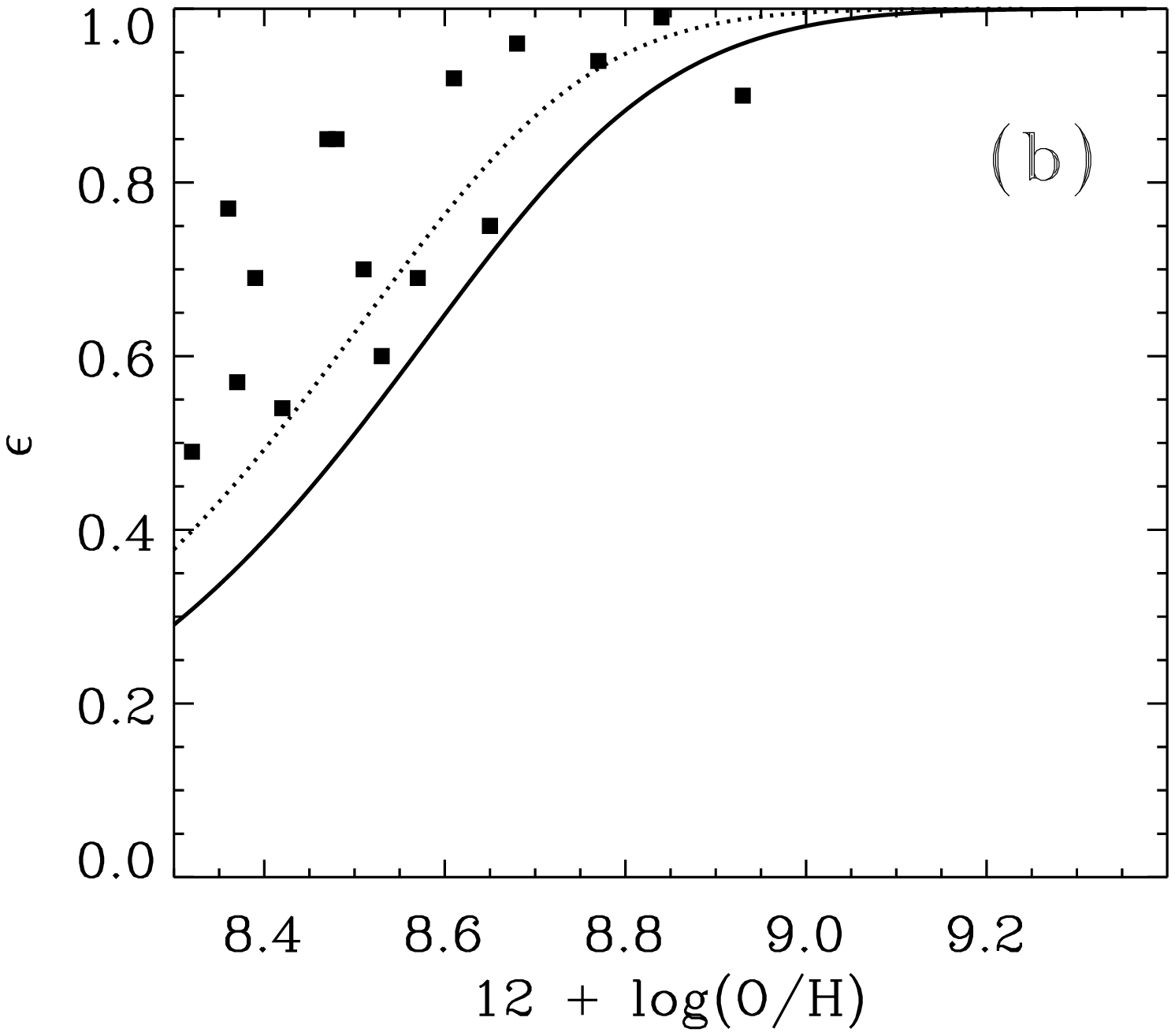}
\end{center}
\caption{Same as Fig.\ \ref{fig:metal_sfg} but for the
{\it IUE}\, sample. The solid line shows the model by H01. The
dotted line represents the result with our new normalisation
for the optical depth of the.}
\label{fig:metal_iue}
\end{figure*}

The correlation between $\epsilon$ and metallicity
(Figure \ref{fig:metal_sfg}b; $r=0.53$) may directly support
H01's idea. H01 postulate a proportionality between the dust
optical depth for the nonionising photons from young
stars ($\tau_{\rm nonion}$) and dust-to-gas mass ratio ${\cal D}$
as
\begin{eqnarray}
\tau_{\rm nonion}=\tau_0\left(\frac{{\cal D}}{10^{-2}}
\right)\, ,
\end{eqnarray}
where they calibrated the normalisation $\tau_0=3.8$ based
on the Galactic condition. Since most of the nonionising photons
are emitted in UV, $\tau_{\rm nonion}$ is related to
$\epsilon$ in a straightforward way:
\begin{eqnarray}
\epsilon =1-{\rm e}^{-\tau_{\rm nonion}}\, .
\end{eqnarray}
By using the conversion from dust-to-gas ratio to metallicity as
depicted in the solid line in H01's Fig.\ 4, we finally obtain
the model relation between metallicity and $\epsilon$. We show
this relation in Figure \ref{fig:metal_sfg}b by the solid line,
which significantly overestimates the observed $\epsilon$. This
implies that the normalisation $\tau_0$ is too large for the
SFG sample.

Then, we lower the normalisation $\tau_0$ to make the model
applicable to the SFG sample. We adopt $\tau_0=1.2$ so that
the mean $\epsilon =0.53$ (i.e., $\tau_{\rm nonion}=0.76$) is
satisfied at the mean metallicity ($12+\log{\rm (O/H)}=8.8$).
The $\epsilon$--metallicity relation under this lower
normalisation is shown by the dotted
line. This reproduce the
observed trend quite well. This implies that H01's model with
$\tau_0=1.2$ can be applicable to star-forming galaxies.

We examine the same relations for the {\it IUE}\, sample in
Figure \ref{fig:metal_iue}. Also for this sample, there is not
any clear trend in the $f$--metallicity diagram ($r=-0.07$), but
there is a correlation between $\epsilon$ and metallicity
($r=0.74$). Thus, we have confirmed the correlation between
metallicity and UV extinction for the {\it IUE}\, sample
(Heckman et al.\ \cite{heckman98}). The solid line in
Figure \ref{fig:metal_iue}b shows the prediction by H01 (i.e.,
$\tau_0=3.8$). Contrary to the SFG
sample, $\tau_0=3.8$ is too small for the {\it IUE}\, sample.
If we assume $\tau_0=5.5$ to satisfy the mean $\epsilon =0.78$
(i.e., $\tau_{\rm nonion}=1.3$) at the mean metallicity
($12+\log{\rm (O/H)}=8.6$), we obtain
the dashed line. However, even in this case, the data points
cannot be reproduced because the extinction is extremely
large even for low-metallicity galaxies in this sample.

Therefore, although H01's idea that there should be a relation
between metallicity and extinction could be partly supported,
the extinction is not described
solely by a function of metallicity as their original idea. The
extinction is
largely dependent on whether a galaxy is a ``starburst''
galaxy or a mild star-forming galaxy. The larger UV optical
depth for the {\it IUE}\, starburst galaxies implies that the
star-forming regions of starburst galaxies are deeply
embedded in dusty gas. Those two classes of galaxies may
also be different in the IR/UV vs.\ UV spectral slope
relation (e.g., Bell \cite{bell02}), which also implies a
fundamental difference in the extinction properties.
Since our method is aimed at a simple treatment to allow
for easy applications, the physical modelling of the
difference is beyond the scope of this paper and is left
for future works.

Moreover, there is an appreciable scatter in the
$\epsilon$--metallicity relation. The scatter can be
caused also by the varieties of following quantities: the
inclination, the geometry of dust distribution, the
dispersion in the relation
between dust-to-gas ratio and metallicity, etc.

It is important that within each type of galaxies, there is a
correlation between extinction and metallicity. This correlation
is equivalent to the correlation between IR/UV flux ratio and
metallicity. It is well known that there is a correlation between
galaxy mass (or luminosity) and metallicity
(Zaritsky et al.\ \cite{zaritsky94};
Richer \& McCall \cite{richer95}; Garnett et al.\ \cite{garnett97}).
There is also a correlation
between mass (or luminosity) and IR/UV flux ratio (or extinction)
(Wang \& Heckman \cite{wang96}; Heckman et al.\ \cite{heckman98};
Buat \& Burgarella \cite{buat98};
Buat et al.\ \cite{buat99}; B02). Those correlations suggest that
large galaxies work as larger reservoirs of gas and metals (and
dust) (Wang \& Heckman \cite{wang96}).

\section{Summary and discussion}\label{sec:summary}

\subsection{Summary}

In this paper, we analysed various SFR indicators (UV, IR, and
H$\alpha$ luminosities). Especially, we focused on the IHK formula
that converts dust IR luminosity into SFR. For this conversion,
the following three quantities are crucial: the fraction of
ionising radiation absorbed by gas ($f$), the fraction of UV
luminosity absorbed by dust ($\epsilon$), and the fraction of
``old'' stellar contribution to the total dust IR luminosity
($\eta$). Those three quantities were observationally estimated
from the 2000 \AA\ monochromatic luminosity, the H$\alpha$
luminosity, and the dust IR luminosity for the SFG sample and the
{\it IUE}\, sample compiled in B02.

The SFG sample proved to have $f=0.57\pm 0.21$,
$\epsilon =0.53\pm 0.21$, and $\eta =0.40\pm 0.06$. Those values
mean that (1) about 40\% of the ionising photons are directly
absorbed by dust; (2) roughly half of the UV photons are
absorbed by dust; (3) about 40\% of the heating of dust is due
to stars older than $10^8$ yr.
For the {\it IUE}\, sample, we found that $f=0.48\pm 0.20$,
$\epsilon =0.76\pm 0.15$, and $\eta =-0.04\pm 0.09$.
Therefore, the typical properties of the {\it IUE}\, sample is as
follows: (1) about 50\% of the ionising photons are absorbed by
dust; (2) most ($\sim 80$\%) of the UV photons are absorbed by
dust; (3) almost all the heating source for dust grains is the
stars younger than $10^8$ yr.

Based on those parameters, we examined the IHK formula.
The SFR derived from this formula agrees almost exactly with the
best-estimate SFR given by the combination of IR and UV
luminosities (eq.\ \ref{eq:realSFR}). This demonstrates the
reliability of IHK's formula over a wide range in SFR
($\sim 0.1$--100 $M_\odot~{\rm yr}^{-1}$). IHK's formula is
different from that of Kennicutt (\cite{kennicutt98b}), where it is
assumed that the dust IR luminosity is equal to the bolometric
luminosity of young stars. This assumption is equivalent
to the case of $f=0$, $\epsilon =1$, and $\eta =0$.
We call this assumption `` dusty starburst approximation''.
For the dusty starburst approximation, our Starburst99
calculation indicates the conversion factor of
$C_{\rm IR}^{\rm sb}=1.79\times 10^{-10}~\solyr~L_\odot^{-1}$. Our
result for the SFG sample implies that
$f\simeq 0.57$, $\epsilon\simeq 0.53$, and $\eta\simeq 0.40$
are applicable for nearby normal star-forming galaxies as a first
approximation, and we obtain the conversion factor
$C_{\rm IR}=2.0\times 10^{-10}~\solyr~L_\odot^{-1}$, a value similar
to that under the dusty starburst approximation. This similarity
comes from the two offsetting effects as stated in
Kennicutt (\cite{kennicutt98a}): the contribution of old stars to
the total IR luminosity and the escape of UV photons without being
absorbed by dust
(Section \ref{subsec:sfr_comp}). The IHK formula works also for the
{\it IUE} starburst sample with $f\simeq 0.48$,
$\epsilon\simeq 0.76$, and $\eta\simeq 0.0$
($C_{\rm IR}=2.4\times 10^{-10}~\solyr~L_\odot^{-1}$).

The SFG sample can be regarded as ``normal'' star-forming galaxies,
and the {\it IUE}\, sample can be representative of the
``starburst'' galaxies. After analysing various SFR indicators, we
found the following (see also Table \ref{tab:formula}):
\begin{itemize}
\item {\bf UV}: The UV SFR should be corrected for dust extinction
by multiplying $1/(1-\epsilon )$. The correction factor depends
largely on the property of individual galaxies, especially on the
starburst/normal category. Among each population (especially among
normal galaxies), $\epsilon$ is systematically small for
${\rm SFR}\la 1~\solyr$.
Thus, extinction estimate for
each galaxy by using e.g., IR luminosity, is important to obtain
a reliable SFR. Panuzzo et al.\ (\cite{panuzzo03})
conclude that the UV luminosity corrected by using the IR/UV ratio
is a reliable indicator of SFR.
\item {\bf H$\vec{\alpha}$}: If the Balmer decrement is measured
precisely enough to correct for the extinction of H$\alpha$
photons, H$\alpha$ luminosity is the most ``secure'' estimator of
SFR,  This is partly because the
correction factor ($1/f$) for the Lyman continuum photons does not
differ between normal and starburst galaxies, and partly because the
there is no systematic trend of $f$ with respect to the SFR. The
dispersion of SFR(H$\alpha$) relative to the SFR(best) (SFR
estimated from UV and IR) can be produced by age variation of the
present star formation activity.
\item {\bf dust IR}: The IR luminosity traces the SFR quite well.
The conversion factor derived under the dusty starburst
approximation is
applicable to both normal and starburst galaxies. There is a risk
that the SFR is underestimated for ${\rm SFR}\la 1~\solyr$. The IHK
formula also provides us with a way to estimate the
conversion factor if we know typical values of $f$, $\epsilon$,
and $\eta$.
\item {\bf Combination of IR and UV}: The simple sum of IR and UV
SFRs systematically overestimates the SFR of normal galaxies,
because some fraction ($\eta$) of IR luminosity is not related to
recent star
formation. If we know the typical $\eta$, we can use equation
(\ref{eq:realSFR}) to subtract the contribution from old stars
to IR luminosity. 
This SFR is free from any SFR-dependent systematics. If we know
$\eta$ for each galaxy, we obtain a reliable estimate of
the SFR.
\end{itemize}

The metallicity dependence of $f$ and $\epsilon$ was also tested.
We found a correlation between $\epsilon$ and metallicity for both
samples, but we did not find any trend of the
$f$--metallicity relation. The $\epsilon$--metallicity relation
of the SFG sample implies lower extinction than that suggested
by H01 ($\tau_{\rm nonion}\simeq 0.76$ on average). On the
contrary, the {\it IUE}\, sample showed a higher extinction by
2.1 times ($\tau_{\rm nonion}\simeq 1.6$ on average). Compared at
the same metallicity level, the {\it IUE}\, sample has the UV
optical depth
4.6 times larger than the SFG sample. This is consistent with
the picture that starburst galaxies are highly obscured by
dust grains (e.g., Heckman et al.\ \cite{heckman98}).

\subsection{Application to the cosmic SFH}\label{subsec:cosmicSFH}

We comment on the application of our method to a cosmological
context. The cosmological evolution of galaxies is one of the
main topics in the cosmic structure formation
(e.g., White \& Rees \cite{white78}). In particular, it
has been an important and unsolved question how and when
galaxies have formed stars
(Tinsley \& Danly \cite{tinsley80}). Such a cosmic SFH as
observationally derived by Madau et al.\ (\cite{madau96})
provides some keys for the statistical view of galaxy evolution.

Takeuchi et al.\ (\cite{takeuchi01}) apply IHK's formula to
derive the cosmic SFH from IR data. They mainly use the number
counts by {\it ISO}\, and the cosmic IR background by {\it COBE}\,
to constrain the comoving IR luminosity evolution. They multiply
$C_{\rm IR}$ (conversion factor of IHK) to the comoving IR
luminosity density and derive the cosmic SFH. They also show that
if there is difference in $f$ and $\epsilon$ between nearby and
distant galaxies, the SFH
derived for the high-$z$ ($z\ga 1$) universe is uncertain by a
factor of $\sim 4$. Therefore, the application of our method to
obtain a typical values for $f$, $\epsilon$, and $\eta$ for
high-$z$ galaxies is an interesting future topic. High-$z$ galaxies
might show a large dust extinction (i.e., small $f$ and large
$\epsilon$) (e.g.,
Heckman et al.\ \cite{heckman98}; Meurer et al.\ \cite{meurer99};
Massarotti et al.\ \cite{massarotti01}) or perhaps a
small extinction (e.g., H01).

The typical properties of a local sample can be applied to
infer the comoving density of SFR at $z=0$. The luminosity of
galaxies per unit comoving volume at $z=0$ has been derived in
a lot of literatures. In  particular, the
2000 \AA\ monochromatic luminosity and the IR
luminosity can be converted to SFR by using the formula
described in this paper. For 2000 \AA\ monochromatic luminosity,
Buat et al.\ (\cite{buat99}) derive the following value for
the comoving density at $z=0$ from the measurement of
Treyer et al.\ (\cite{treyer98}) at $z\sim 0.2$:
\begin{eqnarray}
\rho_{2000}(z=0)=8.9\pm 3.9~10^{37}
h~{\rm erg~s}^{-1}\mbox{\AA}^{-1}~{\rm Mpc}^{-3}\, ,
\end{eqnarray}
where $h$ is the Hubble constant at $z=0$ normalised by
100 km s$^{-1}$ Mpc$^{-1}$. This can be converted to the comoving
SFR by multiplying
$1/(1-\epsilon )$ for extinction correction and $C_{2000}$
for conversion if we know the luminosity-weighted mean of
$1/(1-\epsilon )$ for all the nearby galaxies. We can tentatively
apply the
mean $\epsilon$ derived for the SFG sample ($\sim 0.5$), because
the nearby UV extinction is suggested to be smaller than the
{\it IUE} sample and more similar to that of the SFG sample
(Buat et al.\ \cite{buat99}). Adopting
$\epsilon =0.5$ for the extinction correction, we obtain the
comoving density of SFR at $z=0$ as
\begin{eqnarray}
\rho_{\rm SFR}(z=0)=3.6\pm
1.6~10^{-2}h~\solyr~{\rm Mpc}^{-3}\, .\label{eq:comSFR_UV}
\end{eqnarray}
In fact, we need a complete sample observed in both UV and IR to
estimate dust extinction. Because the available UV samples are
not large enough, we have to wait for future UV observations such
as {\it GALEX}\,\footnote{http://www.srl.caltech.edu/galex/}.

The above $\epsilon (=0.5)$ might be overestimated for a UV
selected sample (K. Xu, private communication). Contrary to it,
an analysis of a currently available UV-selected sample by
Sullivan et al.\ (\cite{sullivan01}) proves a mean UV extinction
to be 1.3 mag, larger than the value which we have adopted above
(0.82 mag). Nevertheless their calculations are made using the
Balmer decrement and the Calzetti extinction curve, and if their
galaxies are similar to the SFG sample they probably
overestimate the extinction (e.g., B02). Thus, future observations
are crucial to correct the comoving UV SFR for dust extinction
even in the local universe.

We can discuss the comoving SFR from IR data.
Saunders et al.\ (\cite{saunders90}) estimate the comoving
density of FIR at $z=0$ (see also
Takeuchi et al.\ \cite{takeuchi03}):
\begin{eqnarray}
\rho_{\rm FIR}(z=0)=5.6\pm 0.6~10^7h~L_\odot~{\rm Mpc}^{-3}\, .
\end{eqnarray}
If we multiply this with 2.4 (the mean value for the SFG sample)
to obtain the total dust IR luminosity, we obtain the comoving
dust IR luminosity as
\begin{eqnarray}
\rho_{\rm IR}(z=0)=1.3\pm 0.1~10^8h~L_\odot~{\rm Mpc}^{-3}\, .
\end{eqnarray}
If we adopt
$C_{\rm IR}(f=0.57,\,\epsilon =0.53,\,\eta =0.40)=2.0\times 10^{-10}$
($\solyr~L_\odot^{-1}$) for the conversion factor from IR
luminosity to SFR (typical for the SFG sample), we obtain the
following local comoving SFR density:
\begin{eqnarray}
\rho_{\rm SFR}(z=0)=2.7\pm 0.3~10^{-2}
h~\solyr~{\rm Mpc}^{-3}\, ,\label{eq:comSFR_IR}
\end{eqnarray}
in good agreement with equation (\ref{eq:comSFR_UV}). An advantage
of using IR luminosity is that we can apply
a similar conversion factor whether a galaxy might be a normal
star-forming one or a starburst one (Table \ref{tab:formula}).
However, we have to be careful about the systematic
underestimate for ${\rm SFR}\la 1~\solyr$. If such
low-SFR galaxies dominate the star formation activity in the
local universe, the above SFR density is an underestimate.

We can make the same kind of argument for the H$\alpha$ comoving
density derived by Gallego et al.\ (\cite{gallego95}) (see also
Tresse et al.\ \cite{tresse02}):
\begin{eqnarray}
\rho_{\rm H\alpha}(z=0)=2.5^{+1.5}_{-0.9}~10^{39}
h~{\rm erg~s^{-1}~Mpc^{-3}}\, .
\end{eqnarray}
In order to convert this to the comoving SFR density, we have to
multiply $C_{\rm H\alpha}/f$ (see equation \ref{eq:formula_ha}).
Then we obtain
$\rho_{\rm SFR}={2.0^{+1.1}_{-0.7}}~{10^{-2}}/
\langle f\rangle~h~\solyr~{\rm Mpc}^{-3}$, where $\langle f\rangle$
is the typical $f$ we should apply to the sample in
Gallego et al.\ (\cite{gallego95}). Since we do not know
$\langle f\rangle$, we assume that it is equal to the mean $f$
(0.57) in the SFG sample. Then, we obtain the following comoving
SFR density:
\begin{eqnarray}
\rho_{\rm SFR}(z=0)=3.5^{+2.0}_{-1.3}~10^{-2}h~\solyr~{\rm Mpc}^{-3}
\, .
\end{eqnarray}
Again we obtain a similar value as the above two estimates
(equations \ref{eq:comSFR_UV} and \ref{eq:comSFR_IR}).
Those comoving SFRs also agree with Buat et al.\ (\cite{buat99}).

We also try to estimate the SFR by using both IR and UV data.
According to equation (\ref{eq:realSFR}), the comoving SFR can be
estimated as
\begin{eqnarray}
\rho_{\rm SFR}(z=0) & = & (1-\eta )C_{\rm IR}^{\rm sb}
\rho_{\rm IR}(z=0)+C_{2000}\rho_{2000}(z=0)\nonumber \\
 & = & 3.2\pm 0.9~10^{-2}h~\solyr~{\rm Mpc}^{-3}\, ,
\end{eqnarray}
where we assume $\eta =0.4$ (the mean value for the SFG sample).
If we apply the simple sum of IR and UV SFRs (i.e., $\eta =0$
in the above estimate), we obtain a larger comoving SFR:
$\rho_{\rm SFR}=(4.1\pm 1.0)~10^{-2}h~\solyr$ Mpc$^{-3}$.
Such a simple sum has been adopted by some authors (e.g.,
Buat et al.\ \cite{buat99}; Flores et al.\ \cite{flores99}),
but this may be an overestimate if a significant fraction of the
dust IR luminosity originates from an old stellar population.

\subsection{Prospects for IR and UV surveys}\label{subsec:prospect}

In the near future, a large IR sample will be obtained by
{\it SIRTF}\,\footnote{http://sirtf.caltech.edu/},
ASTRO-F\,\footnote{http://www.ir.isas.ac.jp/ASTRO-F/index-e.html},
and SOFIA\,\footnote{http://sofia.arc.nasa.gov/} with the typical
redshift $z\la 1$. Even at the first step of the data release, the
galaxy number count can be estimated, to which the modelling by
e.g., Takeuchi et al.\ (\cite{takeuchi01}) can be applied so as to
obtain the comoving density of dust IR luminosity as a function of
$z$ (see also e.g., Gispert et al.\ \cite{gispert00}).
If we only have dust IR luminosity, we can assume a typical values
to apply the IHK formula ($f\sim 0.6$, $\epsilon\sim 0.5$, and
$\eta\sim 0.4$) as a first approximation.

More sensitive and high-resolution IR (or sub-mm) observations by
{\it Herschel}\,\footnote{http://astro.estec.esa.nl/First/},
ALMA\footnote{http://www.eso.org/projects/alma/},
{\it SPICA}\footnote{http://www.ir.isas.ac.jp/SPICA/index.html},
etc.\ will detect a large number of high-$z$ galaxies. For
high-$z$ galaxies, the typical values for $f$, $\epsilon$, and
$\eta$ derived in this paper may not be applicable because of the
difference in dust amount, age, typical size, etc. H01
consider that the
low-metallicity condition at high $z$ makes the dust extinction
less efficient than at low $z$. Then they suggest a high
$C_{\rm IR}$ at high $z$.
Takeuchi et al.\ (\cite{takeuchi01}) also propose that if
$C_{\rm IR}$ is systematically larger at high $z$, we obtain a
flat (or nearly constant) SFH from $z\sim 1$ to $z\sim 5$. On the
contrary, some other works have suggested the importance of dust
extinction for UV photons for high-$z$ galaxies (e.g.,
Meurer et al.\ \cite{meurer99}; Steidel et al.\ \cite{steidel99}).
More recently, Papovich et al.\ (\cite{papovich01}) and
Seibert et al.\ (\cite{seibert02}) have shown that the correction
factor of UV light for dust extinction is $\sim 4$
(i.e., $\epsilon\sim 0.75$) for the Lyman break galaxies at
$z\sim 3$. Therefore, the change of $C_{\rm IR}$ as a function of
$z$ will be an interesting
problem to which we should apply our method.

On the UV side, {\it GALEX}\, data
will be available in a few years. By applying our method to these
data, we will be able to examine the statistical properties of
UV extinction (or $\epsilon$) with the aid of the future IR data.
As mentioned in Section \ref{subsec:cosmicSFH}, this survey will
contribute to revealing the representative value of $\epsilon$ for
the nearby star-forming galaxies with the deepest UV data that
we have ever had. We can also determine the statistical
properties of SFR more accurately by using both UV and IR data
(Flores et al.\ \cite{flores99}; Buat et al.\ \cite{buat99}).

Finally we should comment on galaxies with strong Ly$\alpha$
emission, because we have assumed that all the Ly$\alpha$
photons are absorbed by dust during the resonant scattering.
This assumption may not be valid for objects with large
$L_{\rm UV}/L_{\rm IR}$ because absorption of light by dust is
not efficient in such galaxies.
Such a condition could be satisfied in a primeval galaxies
which is little enriched by dust (or metals). Our assumption is
only valid if
$L_{\rm UV}/L_{\rm IR}<L_{\rm UV}/L_{\rm Ly\alpha}
\simeq 15/f$ (this value comes from Starburst99 prediction).
If dust extinction is inefficient, $f\sim 1$.
Therefore, if we find a galaxy with
$L_{\rm UV}/L_{\rm IR}\ga 15$, it
can be a candidate for a dust-deficient primeval objects
with a conspicuous Ly$\alpha$ emission line.

\begin{acknowledgements}

We are grateful to the anonymous referee for useful comments
which improved this paper very much. We are grateful to
A. Boselli for careful reading and useful
comments and P. Panuzzo for sending us his paper before
publication. We also thank J.-M. Deharveng, D. Burgarella,
J. Iglesias-P\'{a}ramo, B. Milliard, J. Donas, L. Tresse,
T. T. Takeuchi, K. Xu, and A. Ferrara for
stimulating discussions. Two of us (HH and AKI) thank the
hospitality of all
the members at Laboratoire d'Astrophysique de Marseille during
their stay, and the financial support of the Research Fellowship of
the Japan Society for the Promotion of Science
for Young Scientists.  We fully utilised the NASA's Astrophysics Data
System Abstract Service (ADS).

\end{acknowledgements}

%% reference list

\appendix

\section{Conversion factors with a constant SFR over $10^7$ yr}
\label{app:sfr1e7}

In order to see the robustness of the result against
$t_{\rm SF}$, we also construct the conversion formula
for SFR for $t_{\rm SF}=10^7$ yr. In other words, we adopt
the Starburst99 stellar synthetic spectrum at $10^7$ yr with a
constant star formation
history. Since the luminosity of the Lyman continuum photons is
already stationary at $10^7$ yr, the
quantities concerning the Lyman continuum photons and
recombination processes are not changed at all.
We should however change $C_{\rm IR}$ and $C_{2000}$.
If we adopt $t_{\rm SF}=10^7$ yr, the term ``young'' in
should be used to indicate recent $10^7$ yr instead of $10^8$ yr
as in the text. Accordingly ``old'' should be used for the age
larger than $10^7$ yr.

\subsection{$C_{\rm IR}$}

In the main text, we have assumed the stellar synthetic spectrum
made with Starburst99 with a constant SFR at the age of
$10^8$ yr. We examine the Starburst99 result for a constant SFR
at $10^7$ yr with the other conditions fixed. As a result, we
obtain $L_{\rm Lyc}=0.20L_{\rm bol}$ and
$L_{\rm nonion}=0.80L_{\rm bol}$. Since both $L_{\rm Ly\alpha}$
and $L_{\rm ion}$ are already stationary at $10^7$ yr for a
constant SFH, the relation between those two luminosities is
the same as that in the main text:
$L_{\rm Ly\alpha}=0.34fL_{\rm Lyc}$. Instead of equation
(\ref{eq:SF_bol}), we obtain
\begin{eqnarray}
L_{\rm IR}^{\rm SF}=(0.20-0.13f+0.80\epsilon )\,
L_{\rm bol}^{\rm SF}\, .\label{eq:IHKfactor1e7}
\end{eqnarray}
We also obtain the following relation instead of equation
(\ref{eq:SFR_bol}):
\begin{eqnarray}
\frac{\rm SFR}{\solyr}=2.72\times 10^{-10}\,
\frac{L_{\rm bol}^{\rm SF}}{L_\odot}\, .
\end{eqnarray}
The conversion factor (eq.\ \ref{eq:def_cir}) is expressed as
\begin{eqnarray}
C_{\rm IR}(f,\,\epsilon ,\,\eta )=
\frac{2.72\times 10^{-10}(1-\eta )}{0.20-0.13f+0.80\epsilon}~
[\solyr~L_\odot^{-1}]\, .
\end{eqnarray}
The conversion factor under the dusty starburst approximation
(i.e., $f=0$, $\epsilon =1$, and $\eta =0$) becomes
\begin{eqnarray}
C_{\rm IR}^{\rm sb}=2.72\times 10^{-10}~[\solyr~L_\odot^{-1}]\, .
\end{eqnarray}

\subsection{$C_{2000}$}

Because the 2000 \AA\ monochromatic luminosity continues to
increases after $10^7$ yr, $C_{2000}$ at $10^7$ yr is larger than
that at $10^8$ yr. The Starburst99 result indicates that
$C_{2000}=3.18\times 10^{-40}~M_\odot$ yr$^{-1}$ erg$^{-1}$ s \AA .


\begin{thebibliography}{}
\bibitem[1989]{anders89} Anders, E., \& Grevesse, N. 1989, Geochim.\
    Cosmochim.\ Acta, 53, 197
\bibitem[2000]{barger00} Barger, A. J., Cowie, L. L., \& Richards, E.
    2000, AJ, 119, 2092
\bibitem[2002]{bell02} Bell, E. F. 2002, ApJ, 577, 150
\bibitem[2001]{bell01} Bell, E. F., \& Kennicutt, R. C., Jr.\ 2001,
    ApJ, 548, 681
\bibitem[1999]{blain99} Blain, A. W., Smail, I., Ivison, R. J.,
    \& Kneib, J.-P. 1999, MNRAS, 302, 632
\bibitem[2002]{buat02} Buat, V., Boselli, A., Gavazzi, G., \& 
    Bonfanti, C. 2002, A\&A, 383, 801 (B02)
\bibitem[1998]{buat98} Buat, V., \& Burgarella, D. 1998, A\&A, 334, 772
\bibitem[1999]{buat99} Buat, V., Donas, J., Milliard, B., \& Xu, C.
    1999, A\&A, 352, 371
\bibitem[1996]{buat96} Buat, V., \& Xu, C. 1996, A\&A, 306, 61
\bibitem[2001]{calzetti01} Calzetti, D. 2001, PASP, 113, 1449
\bibitem[2000]{calzetti00} Calzetti, D., Armus, L., Bohlin, R. C.,
    Kinney, A. L., Koornneef, J., \& Storchi-Bergmann, T. 2000, ApJ,
    533, 682
\bibitem[1994]{calzetti94} Calzetti, D., Kinney, A. L., \&
    Storchi-Bergmann, T. 1994, ApJ 429, 582
\bibitem[1989]{cardelli89} Cardelli, J. A., Clayton, G. C., \&
    Mathis, J. S. 1989, ApJ, 345, 245
\bibitem[2000]{charlot00} Charlot, S., \& Fall, M. 2000, ApJ, 539,
    718
\bibitem[2002]{charlot02} Charlot, S., Kauffmann, G., Longhetti, M.,
    et al.\ 2002, MNRAS, 330, 876
\bibitem[2001]{charlot01} Charlot, S., \& Longhetti, M. 2001, MNRAS,
    323, 887
%%\bibitem[1992]{condon92} Condon, J. J. 1992, ARA\&A, 30, 575
\bibitem[2000]{cox00} Cox, A. N. 2000, Allen's Astrophysical Quantities,
    4th edn.\ (Springer, New York)
\bibitem[1998]{cram98} Cram, L., Hopkins, A., Mobasher, B., \&
    Rowan-Robinson, M. 1998, ApJ, 507, 155
\bibitem[2002]{dale02} Dale, D. A., \& Helou, G. 2002, ApJ, 576, 159
\bibitem[2001]{dale01} Dale, D. A., Helou, G., Contursi, A.,
    Silbermann, N. A., \& Kolhatkar, S. 2001, ApJ, 549, 215
\bibitem[2001]{deharveng01} Deharveng, J.-M., Buat, V., Le Brun, V.,
    Milliard, B., Kunth, D., Shull, J. M., \& Gry, C. 2001, A\&A,
    375, 805
\bibitem[1994]{deharveng94} Deharveng, J.-M., Sasseen, T. P.,
    Buat, V., Bowyer, S., \& Wu, X. 1994, A\&A, 289, 71
\bibitem[2003]{dopita03} Dopita, M. A., Groves, B. A.,
    Sutherland, R. S., \& Kewley, L. J. 2003, ApJ, 583, 727
\bibitem[1998]{dwek98} Dwek, E. 1998, ApJ, 501, 643
\bibitem[2001]{edmunds01} Edmunds, M. G. 2001, MNRAS, 328, 223
\bibitem[1997]{fioc97} Fioc, M., \& Rocca-Volmerange, B. 1997, A\&A,
    326, 450
\bibitem[2003]{fernandez03} Fern\'{a}ndez-Soto, A., Lanzetta, K. M.,
    \& Chen, H.-W. 2003, MNRAS, in press
\bibitem[1999]{flores99} Flores, H., Hammer, F., Thuan, T. X.,
    et al.\ 1999, ApJ, 517, 148
\bibitem[1989]{gallagher89} Gallagher, J. S., Hunter, D. A., \&
    Bushouse, H. 1989, AJ, 97, 700
\bibitem[1995]{gallego95} Gallego, J., Zamorano, J.,
    Aragon-Salamanca, A., \& Rego, M., 1995, ApJ, 455, L1
\bibitem[1997]{garnett97} Garnett, D. R., Shields, G. A.,
    Skillman, E. D., et al.\ 1997, ApJ, 489, 63
\bibitem[2002]{giallongo02} Giallongo, E., Cristiani, S.,
    D'Odorico, S. \& Fontana, A. 2002, ApJ, 568, L9
\bibitem[2000]{gispert00} Gispert, R., Lagache, G., \& Puget, J.-L.
    2000, A\&A, 360, 1
\bibitem[2001]{hauser01} Hauser, M. G., \& Dwek, E. 1998, ARA\&A, 39,
    249
\bibitem[1998]{heckman98} Heckman, T. M., Robert, C., Leitherer, C.,
    Garnett, D. R., \& van den Rydt, F. 1998, ApJ, 503, 646
\bibitem[2001]{heckman01} Heckman, T. M.,  Sembach, K. R.,
    Meurer, G. R., Leitherer, C., Calzetti, D., Martin, C. L. 2001,
    ApJ, 558, 56
\bibitem[1999]{hiro99} Hirashita, H. 1999, ApJ, 510, L99
\bibitem[2001]{hiro01} Hirashita, H., Inoue, A. K., Kamaya, H., \&
    Shibai, H. 2001, A\&A, 366, 83 (H01)
\bibitem[2002]{hirashita02} Hirashita, H., Tajiri, Y. Y., \&
    Kamaya, H. 2002, A\&A, 388, 439
\bibitem[2001]{hopkins01} Hopkins, A. M., Connolly, A. J.,
    Haarsma, D. B., \& Cram, L. E. 2001, AJ, 122, 288
\bibitem[2001]{inoue01} Inoue, A. K. 2001, AJ, 122, 1788
\bibitem[2002a]{inoue02a} Inoue, A. K. 2002a, ApJ, 570, L97
\bibitem[2002b]{inoue02b} Inoue, A. K. 2002b, ApJ, 570, 688
\bibitem[2000]{inoue00} Inoue, A. K., Hirashita, H., \& Kamaya, H.
    2000, PASJ, 52, 539 (IHK)
\bibitem[2001]{inouehk01} Inoue, A. K., Hirashita, H., \& Kamaya, H.
    2001, ApJ, 555, 613
\bibitem[1990]{issa90} Issa, M. R., MacLaren, I., \& Wolfendale, A. W.
    1990, A\&A, 236, 237
\bibitem[1997]{kamaya97} Kamaya, H., \& Takeuchi, T. T. 1997, PASJ, 49,
    271
\bibitem[1983]{kennicutt83} Kennicutt, R. C., Jr.\ 1983, ApJ, 272, 54
\bibitem[1998a]{kennicutt98a} Kennicutt, R. C., Jr.\ 1998a, ARA\&A, 36, 189
\bibitem[1998b]{kennicutt98b} Kennicutt, R. C., Jr.\ 1998b, ApJ, 498, 541
\bibitem[2002]{kewley02} Kewley, L. J., Geller, M. J., Jansen, R. A., \&
    Dopita, M. A. 2002, AJ, 124, 3135
\bibitem[1996]{kinney96} Kinney, A. L., Calzetti, D., Bohlin, R. C.,
    McQuade, K., Storchi-Bergmann, T., \& Schmitt, H. R. 1996, ApJ, 467,
    38
\bibitem[1995]{leitherer95} Leitherer, C., Ferguson, H. C.,
    Heckman, T. M., \& Lowenthal, J. D. 1995, ApJ, 454, L19
\bibitem[1999]{leitherer99} Leitherer, C., Schaerer, D., Goldader, J. D.,
    et al.\ 1999, ApJS, 123, 3
\bibitem[1998]{lisenfeld98} Lisenfeld, U., \& Ferrara, A. 1998, ApJ,
    496, 145
\bibitem[1987]{lonsdale87} Lonsdale Persson, C. J., \& Helou, G. 1987,
    ApJ, 314, 513
\bibitem[1996]{madau96} Madau, P., Ferguson, H. C., Dickinson, M.,
    Giavalisco, M., Steidel, C. C., \& Fruchter, A. 1996, MNRAS, 283,
    1388
\bibitem[2001]{massarotti01} Massarotti, M., Iovino, A., \& Buzzoni, A.
    2001, ApJ, 559, L105
\bibitem[1999]{meurer99} Meurer, G. R., Heckman, T. M., \& Calzetti, D.
    1999, ApJ, 521, 64
\bibitem[2001]{misiriotis01} Misiriotis, A., Popescu, C. C.,
    Tuffs, R., \& Kylafis, N. D. 2001, A\&A, 372, 775
\bibitem[2002]{nagata02} Nagata, H., Shibai, H., Takeuchi, T. T.,
    \& Onaka, T. 2002, PASJ, 54, 695
\bibitem[1989]{osterbrock89} Osterbrock, D. E. 1989, Astrophysics
    of Gaseous Nebulae and Active Galactic Nuclei (University
    Science Books, California)
\bibitem[2003]{panuzzo03} Panuzzo, P., Bressan, A., Granato, G. L.,
    et al.\ 2003, A\&A, submitted
\bibitem[2001]{papovich01} Papovich, C., Dickinson, M., \&
    Ferguson, H. C. 2001, ApJ, 559, 620
\bibitem[1972]{petrosian72} Petrosian, V., Silk, J., \& Field, G. B.
    1972, ApJ, 177, L69
\bibitem[1995]{richer95} Richer, M. G., \& McCall, M. L. 1995, ApJ,
    445, 642
\bibitem[2002]{rosa02} Rosa-Gonz\'{a}lez,~D., Terlevich,~E., \&
    Terlevich,~R. 2002, MNRAS, 332, 283
\bibitem[1990]{saunders90} Saunders, W., Rowan-Robinson, M.,
    Lawrence, A., et al.\ 1990, MNRAS, 242, 318
\bibitem[1993]{schmidt93} Schmidt, K.-H., \& Boller, T. 1993,
    Astron.\ Nachr., 314, 361
\bibitem[2002]{seibert02} Seibert, M., Heckman, T. M., \&
    Meurer, G. R. 2002, AJ, 124, 46
\bibitem[1978]{smith78} Smith, L. F., Biermann, P., Mezger, P. G.
    1978, A\&A, 66, 65
\bibitem[1978]{spitzer78} Spitzer, L., Jr.\ 1978, Physical
    Processes in the Interstellar Medium (Wiley, New York)
\bibitem[1999]{steidel99} Steidel, C. C., Adelberger, K. L.,
    Giavalisco, M., Dickinson, M., \& Pettini, M. 1999, ApJ, 519, 1
\bibitem[2001]{steidel01} Steidel, C. C., Pettini, M., \&
    Adelberger, K. L. 2001, ApJ, 546, 665
\bibitem[2001]{sullivan01} Sullivan, M., Mobasher, B., Chan, B.,
    Cram, L., Ellis, R., Treyer, M., \& Hopkins, A. 2001, ApJ,
    558, 72
\bibitem[2001]{takeuchi01} Takeuchi, T. T., Ishii, T. T.,
    Hirashita, H., Yoshikawa, K., Matsuhara, H., Kawara, K., \&
    Okuda, H. 2001, PASJ, 53, 37
\bibitem[2003]{takeuchi03} Takeuchi, T. T., Yoshikawa, K., \&
    Ishii, T. T. 2003, ApJ, in press
\bibitem[1980]{tinsley_sola80} Tinsley, B. M. 1980, Findam.\ Cosmic
    Phys., 5, 287
\bibitem[1980]{tinsley80} Tinsley, B. M., \& Danly, L. 1980, ApJ,
    242, 435
\bibitem[2002]{tresse02} Tresse, L., Maddox, S. J.,
    Le F\`{e}vre, O., \& Cuby, J.-G. 2002, MNRAS, 337, 369
\bibitem[1998]{treyer98} Treyer, M. A., Ellis, R. S., Milliard, B.,
    Donas, J., Bridges, T. J. 1998, MNRAS, 300, 303
\bibitem[1996]{walterbos96} Walterbos, R., \& Greenawalt, B. 1996,
    ApJ, 460, 696
\bibitem[1996]{wang96} Wang, B., \& Heckman, T. M. 1996, ApJ, 457,
    645
\bibitem[1978]{white78} White, S. D. M., \& Rees, M. J. 1978, MNRAS,
    183, 341
\bibitem[2000]{witt00} Witt, A. N., \& Gordon, K. D. 2000, ApJ, 528,
    799
\bibitem[1995]{xu95} Xu, C., \& Buat , V. 1995, A\&A, 293, L65
\bibitem[1994]{zaritsky94} Zaritsky, D., Kennicutt, R. C., Jr.,
    \& Huchra, J. 1994, ApJ, 420, 87
\end{thebibliography}
\end{document}